\documentclass{gGAF2e}

\DeclareTextFontCommand\textsfi{\usefont{OT1}{cmss}{m}{sl}}
\DeclareMathAlphabet\mathsfi{OT1}{cmss}{m}{sl}
\DeclareTextFontCommand\textsfb{\usefont{OT1}{cmss}{bx}{n}}
\DeclareMathAlphabet\mathsfb{OT1}{cmss}{bx}{n}
\DeclareTextFontCommand\textsfbi{\usefont{OT1}{cmss}{bx}{sl}}
\DeclareMathAlphabet\mathsfbi{OT1}{cmss}{bx}{sl}

\newcommand\slsa{\mathsfbi{a}}
\newcommand\slsb{\mathsfbi{b}}
\newcommand\sla{\mathsfi{a}}
\newcommand\slb{\mathsfi{b}}

\newcommand\Real{\mbox{Re}}

\newcommand\Ek{\mbox{\textit{E}}}
\renewcommand\Pr{\mbox{\textit{Pr}}}
\newcommand\Pm{\mbox{\textit{Pm}}}
\newcommand\Ra{\mbox{\textit{Ra}}}
\newcommand\Rm{\mbox{\textit{Rm}}}

\renewcommand{\bb}{\mathbf{B}}
\newcommand{\bv}{\mathbf{V}}
\providecommand\bnabla{\boldsymbol{\nabla}}
\newcommand\bOmega{\boldsymbol{\Omega}}
\newcommand\bg{\mathbf{g}}

\renewcommand{\bm}{\mathbf{\overline{B}}}
\newcommand{\Em}{\mathbf{\overline{E}}}
\newcommand{\jm}{\mathbf{\overline{j}}}
\newcommand{\bmT}{\mathbf{\overline{B}}_\mathrm{T}}
\newcommand{\vm}{\mathbf{\overline{V}}}
\newcommand{\bmr}{\overline{B}_r}
\newcommand{\bmt}{\overline{B}_\vartheta}
\newcommand{\bmp}{\overline{B}_\varphi}
\newcommand{\emf}{\boldsymbol{\mathcal{E}}}
\newcommand{\vxb}{\overline{\mathbf{v}\times\mathbf{b}}}
\newcommand{\curl}{\bnabla\times}

\begin{document}

%\title{Comparing mean-field models with direct numerical simulations of rotating magnetoconvection and the geodynamo}
\title{Mean-field concept and direct numerical simulations of rotating magnetoconvection and the geodynamo}
\author{MARTIN SCHRINNER$\dagger$, KARL-HEINZ R\"ADLER$\ddagger$, DIETER SCHMITT$\dagger$,\\MATTHIAS RHEINHARDT$\ddagger$ and ULRICH R. CHRISTENSEN$\dagger$ \\[2mm]
$\dagger$Max-Planck-Institut f\"ur Sonnensystemforschung, 37191 Katlenburg-Lindau, Germany \\
$\ddagger$Astrophysikalisches Institut Potsdam, 14482 Potsdam, Germany}
\markboth{M. Schrinner, K.-H. R\"adler, D. Schmitt, M. Rheinhardt and U. R. Christensen}{Mean-field concept and direct numerical simulations of rotating magnetoconvection and the geodynamo}
\maketitle

\begin{abstract}
Mean-field theory describes magnetohydrodynamic processes leading to large-scale magnetic fields in
various cosmic objects. In this study magnetoconvection and dynamo processes in a rotating
spherical shell are considered. Mean fields are defined by azimuthal averaging. In the framework of
mean-field theory, the coefficients which determine the traditional representation of the mean
electromotive force, including derivatives of the mean magnetic field up to the first order, are
crucial for analyzing and simulating dynamo action. Two methods are developed to extract mean-field
coefficients from direct numerical simulations of the mentioned processes. While the first method
does not use intrinsic approximations, the second one is based on the second-order correlation
approximation. There is satisfying agreement of the results of both methods for sufficiently slow
fluid motions. Both methods are applied to simulations of rotating magnetoconvection and a
quasi-stationary geodynamo. The mean-field induction effects described by these coefficients, e.g.
the $\alpha$-effect, are highly anisotropic in both examples. An $\alpha^2$-mechanism is suggested
along with a strong $\gamma$-effect operating outside the inner core tangent cylinder. The
turbulent diffusivity exceeds the molecular one by at least one order of magnitude in the geodynamo
example. With the aim to compare mean-field simulations with corresponding direct numerical
simulations, a two-dimensional mean-field model involving all previously determined mean-field
coefficients was constructed. Various tests with different sets of mean-field coefficients reveal
their action and significance. In the magnetoconvection and geodynamo examples considered here, the
match between direct numerical simulations and mean-field simulations is only satisfying if a large
number of mean-field coefficients are involved. In the magnetoconvection example, the azimuthally
averaged magnetic field resulting from the numerical simulation is in good agreement with its
counterpart in the mean-field model. However, this match is not completely satisfactory in the
geodynamo case anymore. Here the traditional representation of the mean electromotive force
ignoring higher than first-order spatial derivatives of the mean magnetic field is no longer a good
approximation.
\\[2mm]
\end{abstract}

\section{Introduction}
\label{sec:1}

Large-scale cosmic magnetic fields as the Earth's, the solar and the galactic magnetic fields are
maintained by hydromagnetic dynamos \cite[e.g.][]{weiss_02}. However, global computational dynamo
models simulate only the geodynamo resonably well, while it remains difficult to tackle the stellar
or galactic dynamo problem in this way. Despite the increasing computational power, a direct
numerical treatment of the governing equations is not yet feasible in the latter cases, because of
the huge range of spatial and temporal scales needed to be resolved there
\citep{tobias_02,weiss_00,shukurov_02}. First attempts of global MHD simulations have been carried
out by Gilman \citep{gilman_81,gilman_83} adopting the Boussinesq approximation and later by
Glatzmaier \citep{glatzmaier_84,glatzmaier_85} using an anelastic model. Both aimed at modelling
the solar dynamo. Though cyclic dynamo solutions were obtained in some cases, the models exhibited
a wrong poleward migration of the magnetic field \citep{glatzmaier_85}. In a recent attempt,
\citet{brun_04} succeeded in simulating a solar-like differential rotation, but could not reproduce
basic features of the solar cycle yet.

An alternative approach is provided by mean-field electrodynamics
\citep{steenbeck_66,moffatt_78,raedler_80}, which is a theory focussing only on large scale, i.e.
averaged fields. Highly complex small-scale or residual parts need not to be known in detail, only
the averaged cross product of the residual velocity and magnetic field, in the following called the
mean electromotive force, is relevant and accounts for the evolution of the mean field.
Advantageously, the difficulties in resolving the small-scale structures can be avoided. Usually,
the action of the small-scale velocity on the mean magnetic field as expressed by the mean
electromotive force is parametrised, and the parameters are the mean-field coefficients defined
below. Most prominent among them are those which describe the $\alpha$-effect. They are closely
related to the fundamental induction effect of cyclonic convection \citep{parker_55}. Other
mean-field coefficients describe the advection and diffusion of the mean magnetic field. In other
words, mean-field theory supplies theoretical insight as well as formalised physical concepts in
order to interpret and, in principle, also to quantify dynamo action.

Despite their relative simplicity, mean-field models have successfully reproduced basic features of
the solar cycle \citep[see e.g.][]{steenbeck_69,ossen_03} and are moreover unique in coherently
simulating many features of the magnetic field in spiral galaxies \citep{beck_96,shukurov_02}. But,
whether or not mean-field models show dynamo action depends strongly on the mean-field
coefficients. They are not known a priori but instead in general determined in a heuristic way.

Concerning the geodynamo, the situation is different. Many features of the Earth's magnetic field
are successfully reproduced by nonlinear three-dimensional simulations of the magnetohydrodynamics
in the Earth's core.  Although some model parameters still do not reach realistic values and in
particular viscous effects and therefore the  size of viscous boundary layers are by orders of
magnitude overestimated, the simulations exhibit a magnetic field dominated by an axial dipole at
the Earth's surface that is maintained over several magnetic diffusion times
\citep{glatz_95b,kuang_97,christensen_98}. In addition, the time-dependence of the dipole moment,
including secular variations, excursions and reversals, resembles the observed Earth's magnetic
field \citep{glatz_95a,kutzner_02,wicht_04}. The success in simulating the geodynamo can be
attributed to the rather moderate vigour of turbulence in the Earth's outer core compared to the
much more turbulent dynamics in the solar convection zone. The difference is formally expressed in
terms of very different magnetic Reynolds numbers: While $\Rm\approx 10^3$ for the Earth's outer
core \citep{fearn_98}, a representative value of $\Rm$ at the base of the solar convection zone is
$10^{10}$ \citep{ossen_03}.

Even though present geodynamo models are fully self-consistent, the interpretation of the mechanism
by which they generate magnetic field relies frequently in a heuristic manner on mean-field
concepts \citep{glatz_95b,kageyama_97,olson_99}. Thus mean-field theory is very useful and
indispensable at the present time. However, the applicability of mean-field concepts as tools for
analysing dynamo processes in numerical simulations suffers from the poor knowledge of the
mean-field coefficients and reliable methods to derive them.

There are two seminal approaches which have been followed in order to determine mean-field
coefficients. The first one aims at deriving (quasi-)analytical expressions of the mean
electromotive force and the mean-field coefficients \citep[see e.g.][]{moffatt_78,raedler_80}. This
requires the integration of the governing equation for the residual magnetic field with the help of
closure methods. Most commonly used is the second-order correlation approximation (SOCA) in which
only statistical moments up to the second order are taken into account, while moments of higher
order are neglected. Early investigations with SOCA and simple assumptions on the turbulence,
starting with the seminal paper by \citet{steenbeck_66}, are summarized in
\citet{raedler_80b,raedler95,raedler00c}. More sophisticated assumptions on the turbulent motion
let \citet{kichatinov_92} and \citet{ruediger_93} provide $\alpha$-coefficients in the high
conductivity limit for various rotation rates. In the context of the galactic dynamo,
\citet{ferriere_92,ferriere_93a,ferriere_93b} considered supernova explosions and derived
expressions for the components of the $\boldsymbol{\alpha}$- and $\boldsymbol{\beta}$-tensors in
cylindrical geometry. Steps beyond SOCA for specific flows were undertaken by
\citet{nicklausetal88}, \citet{carvalho92,carvalho94} and \citet{schmitt84,schmitt03}. In the
context of the Karlsruhe dynamo experiment the velocity field possesses a number of symmetries,
which allow a rather direct computation of mean-field coefficients
\citep{raedleretal97,raedleretal98,raedleretal02,raedleretal02b}. A particular approach beyond SOCA
is the so-called $\tau$-approximation, first introduced by \citet{vainshteinetal83}, recently
revived by \citet{rogachevskiietal03}, \citet{raedleretal03} and
\citet{brandenburgetal05,brandenburgetal05b}, and critically reviewed by \citet{raedleretal06d}.

The second approach to the mean-field coefficients makes use of numerical modelling, in which the
mean electromotive force, $\emf$, as well as the mean-field, $\bm$, are determined numerically as
output of a MHD-simulation. Further on, a linear relation between $\emf$ and $\bm$ is assumed,
which has to be inverted in order to solve for the unknown mean-field coefficients. A fundamental
problem related to this approach is that the number of unknown variables is in general much higher
than the number of equations resulting from the linear relation between $\emf$ and $\bm$.
Therefore, all work presented so far refers either to specific situations in which certain
constraints reduce the number of mean-field coefficients beforehand, or some of the mean-field
coefficients are considered as small and are neglected. \citet{ziegler_96} calculated the
$\boldsymbol{\alpha}$-tensor for the galactic dynamo on the basis of numerical simulations of
supernova explosions and confirmed the results given by \citet{ferriere_93a}. Following a similar
approach, \citet{ossendrijver_01,ossendrijver_02} and \citet{kaepylaeetal06} considered
magnetoconvection in the solar convection zone and used box simulations to determine the
$\alpha$-tensor in Cartesian geometry.
%Both, \citet{ziegler_96} and \citet{ossendrijver_02} performed three numerical
%experiments with imposed homogeneous mean magnetic field in orthogonal directions and combined the
%results in order to close the system of linear equations.
Similarly, \citet{gieseckeetal05} derived a geodynamo $\alpha$-effect from box simulations of
rotating magnetoconvection. In a further attempt to determine not only the $\boldsymbol{\alpha}$-
but the $\boldsymbol{\beta}$-tensor as well for accretion and galactic disks,
\citet{brandenburg_02} and \citet{kowaletal06} applied box simulations of turbulence.
\citet{schrinner05} calculated all mean-field coefficients in a spherical shell with a convection
pattern relevant for the geodynamo. The method and results are described in detail in this paper.
For shorter contributions see also \citet{schrinneretal05,schrinneretal06}.

The idea of this work is to take advantage of global numerical simulations of rotating
magnetoconvection and the geodynamo and to compare them with respective mean-field calculations,
where mean fields are defined by azimuthal averaging. This will lead to an estimation of the
reliability of mean-field theory and its often used approximations. Furthermore, such a comparison
will help to improve the conceptual understanding of dynamo mechanisms which are observed in the
numerical simulations.

As already pointed out, both aims are intimately associated with the derivation of the
corresponding mean-field coefficients. Hence, emphasis is placed on the developement of two methods
which contribute to each of the principal approches mentioned above. Both methods have been applied
to a simulation of rotating magnetoconvection and a quasi-stationary geodynamo. They are consistent
with each other in a parameter regime in which the second-order correlation approximation is
justified and serve as powerful tools to determine a number of relevant mean-field coefficients.
While most of the quoted earlier work refers to a Cartesian  geometry, global mean-field
coefficients for the astrophysically more relevant domain of a rotating spherical shell are
presented here, and specific problems related to the spherical geometry are discussed.

The plan of the paper is as follows: In section 2 the equations, the boundary conditions and the
parameters of the considered numerical models are given. Section 3 summarizes the mean-field
concept and introduces the mean electromotive force. In section 4 two approaches to determine the
mean-field coefficients are developed, a general numerical one and a semi-analytical approach using
SOCA. In section 5 the mean-field coefficients for two models, rotating magnetoconvection and a
quasi-steady geodynamo, are derived and their quenching as well as limitations of the validity of
SOCA are discussed. A comparison of the mean fields for both models, derived from numerical
simulations and from mean-field theory, respectively, is made in section 6, and the range of
validity of the representation of the mean electromotive force are discussed. Section 7 summarizes
our conclusions. In the appendices details of the derivation of the mean-field coefficients are
given and the mean-field energy balance is considered.

\section{The numerical models considered}
\label{sec:2}

In both models, magnetoconvection and geodynamo, a rotating spherical shell of electrically
conducting fluid is considered in which the fluid velocity $\bv$, the magnetic field $\bb$ and the
temperature $T$ are governed by the following equations using the Boussinesq approximation:
\begin{eqnarray}
& \displaystyle{\frac{\partial\bv}{\partial t}+(\bv\cdot\bnabla)\bv-\nu\nabla^2\bv
+2\,\bOmega\times\bv = -\frac{1}{\varrho}\bnabla P+\alpha_T\bg T +
\frac{1}{\mu\varrho}(\bnabla\times\bb)\times\bb} & \label{eq2:2} \\
& \displaystyle{\frac{\partial\bb}{\partial t}=\bnabla\times(\bv\times\bb) +\eta\nabla^2\bb} &
\label{eq2:8} \\
& \displaystyle{\frac{\partial T}{\partial t}+\bv\cdot\bnabla T  = \kappa\Delta T} & \label{eq2:6} \\
& \displaystyle{\bnabla\cdot\bv=\bnabla\cdot\bb=0} \,. & \label{eq2:10}
\end{eqnarray}
Here $P$ means a modified pressure, $\varrho$ is the mass density of the fluid and $\mu$ its
magnetic permeability. Further $\nu$, $\eta$ and $\kappa$ are kinematic viscosity, magnetic
diffusivity and thermal conductivity, and $\alpha_T$ the thermal expansion coefficient, all
considered as constants. The motion is measured relative to the uniform rotation of the shell with
angular velocity $\bOmega=\Omega\mathbf{e}_z$ where $\mathbf{e}_z$ is the unit vector in the
direction of the rotation axis. The gravitational acceleration is specified by
$\bg=g_0\mathbf{{r}}/r_0$, with $g_0$ being its value at the outer boundary $r=r_0$. The ratio of
inner to outer radius of the shell is $r_i/r_0=0.35$ and thus the shell width $D=0.65\,r_0$ for all
simulations considered here. The original form of the buoyancy term is $\alpha_T\bg(T-T_0)$ with
$T_0$ describing the temperature distribution of a reference state. Here $T_0$ is assumed to depend
on $r$ only. Then $\alpha_T\bg T_0$ can be represented as a gradient, which is absorbed in the
pressure term in equation~(\ref{eq2:2}).

For the velocity $\bv$ no-slip boundary conditions are adopted, $\bv=\mathbf{0}$ at $r=r_i$ and
$r=r_0$. Moreover, all surroundings of the spherical shell are assumed as electrically
non-conducting, so that the magnetic field $\bb$ continues as a potential field in both parts
exterior to the fluid shell. In the magnetoconvection case a toroidal field is imposed via
inhomogeneous boundary conditions. The temperature is assumed to be constant on the boundaries such
that $T=\delta T$ at $r=r_i$ and $T=0$ at $r=r_0$.

Measuring length, time, magnetic field and temperature in units of  $D$, $D^2/\nu$,
$(\varrho\mu\eta\Omega)^{1/2}$ and $\delta T$, the above equations can be written in a
non-dimensional form which contains only four non-dimensional parameters
\citep[e.g.,][]{christensen_01}. These are the Ekman number $\Ek$, the modified Rayleigh number
$\Ra$, the Prandtl number $\Pr$, and the magnetic Prandtl number $\Pm$,
\begin{eqnarray}
&\displaystyle{ \Ek=\nu/\Omega D^2 \,, \quad \Ra=\alpha_T g_0\delta T D/\nu\Omega }& \nonumber\\
&\displaystyle{ \Pr=\nu/\kappa \,, \quad \Pm=\nu/\eta \,.}&
\end{eqnarray}
In order to characterise the results of the simulations, the magnetic Reynolds number $\Rm$ and the
Elsasser number $\Lambda$
\begin{equation}
\Rm=vD/\eta \,,\quad \Lambda=B^2/\varrho\mu\eta\Omega
\end{equation}
with $v$ interpreted as r.m.s. velocity and $B$ as the r.m.s. value of the magnetic field inside
the shell are used.

For the numerical solution of the above equations, a code is used which was constructed in its
original form by \citet{glatzmaier_84}. This version of the code solved the anelastic
magnetohydrodynamic equations in a spherical shell to simulate stellar dynamos.
\citet{olson:glatz95} and \citet{christensen_99} later applied a modified version of the numerical
model to run magnetoconvection and dynamo simulations in a rotating spherical shell adopting the
Boussinesq approximation. The code has been validated by benchmarking it with other
three-dimensional models \citep{christensen_01}.

\section{The mean-field concept}
\label{sec:3}

\subsection{The mean electromotive force and mean-field coefficients}
\label{sec:3:1}

Within the scope of this work, the mean-field concept is applied to the induction equation
(\ref{eq2:8}) only. In the following, we refer to a spherical coordinate system
$(r,\vartheta,\varphi)$ whose polar axis coincides with the rotation axis of the shell. Mean vector
fields are defined by averaging the components of the original fields over all values of the
azimuthal coordinate $\varphi$, e.g., $\bm=\bmr(r,\vartheta)\mathbf{e}_r+
\bmt(r,\vartheta)\mathbf{e}_\vartheta+ \bmp(r,\vartheta)\mathbf{e}_\varphi$ in which
$\bmr(r,\vartheta), \bmt(r,\vartheta)$, and $\bmp(r,\vartheta)$ are the azimuthal averages of
$B_r(r,\vartheta,\varphi)$, $B_\vartheta(r,\vartheta,\varphi)$ and
$B_\varphi(r,\vartheta,\varphi)$. Note that with this definition of mean fields the Reynolds
averaging rules apply exactly. Of course, all mean fields are axisymmetric about the polar axis.

Subjecting the induction equation  (\ref{eq2:8}) to this averaging yields
\begin{equation}
\frac{\partial\bm}{\partial t}=\curl(\vm\times\bm)+\curl\emf+\eta\nabla^2\bm,
\label{eq3:2}
\end{equation}
with the mean electromotive force
\begin{equation}
\emf=\vxb \, ,
\label{eq3:4}
\end{equation}
in which $\mathbf{v}$ and $\mathbf{b}$ are the residual velocity and magnetic fields,
$\mathbf{b}=\mathbf{B}-\bm$ and $\mathbf{v}=\mathbf{V}-\vm$. If $\mathbf{v}$
and $\vm$ are given, the calculation of $\emf$ requires the knowledge of $\mathbf{b}$, which is
governed by
\begin{equation}
\frac{\partial\mathbf{b}}{\partial t}=
\curl(\vm\times\mathbf{b})+\curl({\mathbf{v}\times\bm})+
\curl\mathbf{G}+\eta\nabla^2\mathbf{b} \, ,
\label{eq3:8}
\end{equation}
with $\mathbf{G}=\mathbf{v}\times\mathbf{b}-\vxb$. According to (\ref{eq3:4}) and
(\ref{eq3:8}), $\emf$ is a functional of $\mathbf{v}$, $\vm$, and $\bm$, which is linear in
$\bm$. We adopt the frequently used assumption that $\mathbf{b}$ vanishes if $\bm$ does so.
This excludes the possibility of a dynamo with $\bm=0$, in other context referred to as
``small-scale dynamo". Then $\emf$ is not only linear but also homogeneous in $\bm$ and can be
expressed in the form
\begin{equation}
{\mathcal{E}}_\kappa (r, \vartheta, t) = \int_\mathcal{V} \int_{t'< \, t}
    K_{\kappa\lambda} (r, \vartheta, t; r', \vartheta', t')
    {\overline{B}}_\lambda (r', \vartheta',t') \mbox{d} v' \mbox{d} t'
\label{eq3:10}
\end{equation}
with some kernel $K_{\kappa\lambda}$. Here and in what follows indices like $\kappa$ or $\lambda$
are used for $r, \vartheta$ or $\varphi$, and the summation convention is adopted. The integration
is over the whole fluid shell, $\mathcal{V}$, and all times $t'<\, t$.

It seems plausible to assume that $\emf$ at a given point in space and time depends only on
quantities in certain surroundings of this point. This implies that $K_{\kappa \lambda}$ differs
only for sufficiently small $|r - r'|$, $|\vartheta - \vartheta'|$ and $t - t'$ markedly from zero.
We adopt here the assumption that $\bm$ varies only weakly in space and time so that its behavior
in the relevant surroundings of a given  point can be well described by $\bm$ and its first spatial
derivatives in this point. This brings us from (\ref{eq3:10}) to
\begin{equation}
\mathcal{E}_\kappa=\tilde{a}_{\kappa\lambda}\overline{B}_\lambda
    + \tilde{b}_{\kappa\lambda r}\frac{\partial\overline{B}_\lambda}{\partial r}
    + \tilde{b}_{\kappa\lambda\vartheta}\frac{1}{r}\frac{\partial\overline{B}_\lambda}{\partial\vartheta} \,.
\label{eq3:12}
\end{equation}
Note that all higher than first-order spatial derivatives and all time derivatives of $\bm$ are
ignored. It remains to be checked to which extent their neglect is justified for the considered
examples. The representation (\ref{eq3:12}) contains 27 independent coefficients
$\tilde{a}_{\kappa\lambda}$, $\tilde{b}_{\kappa\lambda r}$ and
$\tilde{b}_{\kappa\lambda\vartheta}$, which are determined by $\mathbf{v}$ and $\vm$ and,
considered as functionals of these quantities, independent of $\bm$. We have
\begin{equation}
\tilde{a}_{\kappa \lambda} (r, \vartheta, t) = \int_\mathcal{V} \int_{t'< \, t}
    K_{\kappa\lambda} (r, \vartheta, t; r', \vartheta', t') \, \mbox{d} v' \mbox{d} t' \, .
\label{eq3:14}
\end{equation}
and analogous relations for $\tilde{b}_{\kappa \lambda r}$ and $\tilde{b}_{\kappa \lambda
\vartheta}$ with additional factors $(r' - r)$ or $r (\vartheta' - \vartheta)$, respectively, in
the integrand.

\subsection{A more general representation of the mean electromotive force}
\label{sec:3:2}

We have introduced the representation (\ref{eq3:12}) of the mean electromotive force $\emf$ under
special conditions. In particular we relied on a spherical coordinate system and restricted
ourselves to mean magnetic fields $\bm$ which are axisymmetric about the polar axis of this system.

In general, the mean electromotive force $\emf$, if no higher than first-order spatial derivatives
and no time derivatives of $\bm$ are taken into account, is written in the form
\begin{equation}
\emf=\slsa \, \bm + \slsb \, \bnabla\bm
\label{eq5:2}
\end{equation}
with a second-rank tensor $\slsa$ and a third-rank tensor $\slsb$. This relation is understood as a
coordinate-independent connection between the vector $\emf$, the vector $\bm$ and its gradient
tensor $\bnabla\bm$. It applies independently of symmetries of $\bm$. In a Cartesian coordinate
system (\ref{eq5:2}) takes the form ${\mathcal{E}}_i = \sla_{ij} {\overline{B}}_j + \slb_{ijk}
\partial {\overline{B}}_j / \partial x_k $. When changing to a curvilinear coordinate system
$\partial {\overline{B}}_j / \partial x_k$ turns into a covariant derivative.

It is useful to rewrite (\ref{eq5:2}) into the equivalent relation
\begin{equation}
\emf = - \boldsymbol{\alpha} \cdot \bm - \boldsymbol{\gamma}\times\bm
    - \boldsymbol{\beta} \cdot (\curl\bm) - \boldsymbol{\delta}\times(\curl\bm)
    - \boldsymbol{\kappa} \cdot (\bnabla\bm)^{(\rm{sym})},
\label{eq5:4}
\end{equation}
see \cite{raedler_80b,raedler00c} or \cite{raedleretal06}. Here $\boldsymbol{\alpha}$ and
$\boldsymbol{\beta}$ are symmetric second-rank tensors, $\boldsymbol{\gamma}$ and
$\boldsymbol{\delta}$ vectors, $\boldsymbol{\kappa}$ is a third-rank tensor symmetric in the
indices connecting it with $(\bnabla\bm)^{(\rm{sym})}$, the latter being the symmetric part of the
gradient tensor $\bnabla \bm$. The relationship between the components of $\slsa$ and $\slsb$ and
those of $\boldsymbol{\alpha}$, $\boldsymbol{\gamma}$, $\boldsymbol{\beta}$, $\boldsymbol{\delta}$
and $\boldsymbol{\kappa}$ is given below.

The representation (\ref{eq5:4}) of $\emf$ allows a discussion of the individual induction effects.
The $\alpha$ term describes the $\alpha$-effect, which is in general anisotropic. The $\gamma$ term
corresponds to a transport of mean magnetic flux like that by a mean motion of the fluid. The
$\beta$ term and also the $\delta$ term can be interpreted by introducing an anisotropic mean-field
conductivity. The $\kappa$ term covers various other influences on the mean field, which are more
difficult to interpret.

In contrast to the 27 coefficients $\tilde{a}_{\kappa\lambda}$, $\tilde{b}_{\kappa\lambda r}$ and
$\tilde{b}_{\kappa\lambda\vartheta}$ in (\ref{eq3:12}) we have here in general 36 independent
components of $\slsa$ and $\slsb$, or of $\boldsymbol{\alpha}$, $\boldsymbol{\beta}$,
$\boldsymbol{\gamma}$, $\boldsymbol{\delta}$ and $\boldsymbol{\kappa}$. The lower number in the
case of (\ref{eq3:12}) is due to the assumed axisymmetry of $\bm$. We stress that the
$\tilde{a}_{\kappa\lambda}$, $\tilde{b}_{\kappa\lambda r}$ and $\tilde{b}_{\kappa\lambda\vartheta}$
should not be considered as tensor components, whereas $\slsa$ and $\slsb$ as well as
$\boldsymbol{\alpha}$, $\boldsymbol{\beta}$, $\boldsymbol{\gamma}$, $\boldsymbol{\delta}$ and
$\boldsymbol{\kappa}$ are indeed tensors or vectors with the usual defining transformation
properties under changes of the coordinate system.

We may, of course specify the relation (\ref{eq5:2}) to our spherical coordinate system and to
axisymmetric $\bm$. When doing so in the above sense, that is, taking the covariant forms of the
derivatives of $\bm$, we arrive at relations for $\mathcal{E}_\kappa$ of the form (\ref{eq3:12})
with
\begin{eqnarray}
& \displaystyle{\tilde a_{\kappa r} = \sla_{\kappa r} + \slb_{\kappa\vartheta\vartheta}/r + \slb_{\kappa\varphi\varphi}/r
\, , \quad \tilde a_{\kappa\vartheta} = \sla_{\kappa\vartheta} - \slb_{\kappa r\vartheta}/r + \cot\vartheta \,
\slb_{\kappa\varphi\varphi}/r \, ,} &
\nonumber\\
& \displaystyle{\tilde a_{\kappa\varphi} = \sla_{\kappa\varphi} - (\slb_{\kappa r\varphi}+\cot\vartheta \,
\slb_{\kappa\vartheta\varphi})/r \, , \quad
\tilde b_{\kappa\lambda r} = \slb_{\kappa\lambda r} \, , \quad
\tilde b_{\kappa\lambda\vartheta} = \slb_{\kappa\lambda\vartheta} \, .} &
\label{eq5:5}
\end{eqnarray}
We may understand (\ref{eq5:5}) as a system of 27 equations which determine the $\sla_{\kappa r} +
\slb_{\kappa \varphi \varphi}/r$, $\sla_{\kappa \vartheta} + \cot \vartheta \, \slb_{\kappa \varphi
\varphi}/r$, $\sla_{\kappa \varphi} - (\slb_{\kappa r \varphi} + \cot \vartheta \slb_{\kappa \vartheta
\varphi})/r$, $\slb_{\kappa\lambda r}$ and $\slb_{\kappa\lambda\vartheta}$ if the $\tilde a_{\kappa
\lambda}$, $\tilde b_{\kappa \lambda r}$ and $\tilde b_{\kappa \lambda \vartheta}$ are known.
Clearly we have the freedom of arbitrarily choosing the $\slb_{\kappa \lambda \varphi}$. Of course,
this choice also influences the $\sla_{\kappa \lambda}$. It is, however, without any influence on
$\emf$. For what follows we put simply $\slb_{\kappa\lambda\varphi}=0$. Then we have
\begin{eqnarray}
& \displaystyle{\sla_{\kappa r} = \tilde a_{\kappa r}-\tilde b_{\kappa\vartheta\vartheta}/r \, , \quad
\sla_{\kappa\vartheta} = \tilde a_{\kappa\vartheta}+\tilde b_{\kappa r\vartheta}/r \, , \quad
\sla_{\kappa\varphi} = \tilde a_{\kappa\varphi}} &
\nonumber\\
& \displaystyle{\slb_{\kappa\lambda r} = \tilde b_{\kappa\lambda r} \, , \quad
\slb_{\kappa\lambda\vartheta} = \tilde b_{\kappa\lambda\vartheta} \, ,  \quad
\slb_{\kappa \lambda \varphi} = 0 \, .} &
\label{eq5:6}
\end{eqnarray}

Using this we may also express the components of the $\boldsymbol{\alpha}$, $\boldsymbol{\beta}$,
$\boldsymbol{\gamma}$, $\boldsymbol{\delta}$ and $\boldsymbol{\kappa}$ in the spherical coordinate
system by the $\tilde a_{\kappa \lambda}$, $\tilde b_{\kappa \lambda r}$ and $\tilde b_{\kappa
\lambda \vartheta}$. The relations of the components of $\boldsymbol{\alpha}$,
$\boldsymbol{\gamma}$, $\boldsymbol{\beta}$, $\boldsymbol{\delta}$ and $\boldsymbol{\kappa}$ to
that of $\slsa$ and $\slsb$ are given by
\begin{eqnarray}
&\displaystyle{\alpha_{\kappa\lambda} = -1/2(\sla_{\kappa\lambda}+\sla_{\lambda\kappa}) \,, \quad
\gamma_\kappa = 1/2\epsilon_{\kappa\lambda\mu}\sla_{\lambda\mu}}& \nonumber\\
&\displaystyle{\beta_{\kappa\lambda} = 1/4(\epsilon_{\kappa\mu\nu}\slb_{\lambda\mu\nu}+\epsilon_{\lambda\mu\nu}\slb_{\kappa\mu\nu})}& \\
&\displaystyle{\delta_\kappa = -1/4(\slb_{\lambda\kappa\lambda}-\slb_{\lambda\lambda\kappa}) \,, \quad
\kappa_{\kappa\lambda\mu} = -1/2(\slb_{\kappa\lambda\mu}+\slb_{\kappa\mu\lambda})} \,.& \nonumber
\end{eqnarray}
Combining this with (\ref{eq5:6}) we find
\begin{eqnarray}
&\displaystyle{\alpha_{rr} =  -(\tilde{a}_{rr}-\tilde{b}_{r\vartheta\vartheta}/r)
\,, \quad
\alpha_{r\vartheta}  =  \alpha_{\vartheta r}=-1/2\,(\tilde{a}_{r\vartheta}+
\tilde{b}_{rr\vartheta}/r+\tilde{a}_{\vartheta r}-
\tilde{b}_{\vartheta\vartheta\vartheta}/r)}&
\nonumber\\
&\displaystyle{\alpha_{r\varphi} = \alpha_{\varphi r}=-1/2\,(\tilde{a}_{r\varphi}
+\tilde{a}_{\varphi r}-\tilde{b}_{\varphi\vartheta\vartheta}/r)
\,, \quad
\alpha_{\vartheta\vartheta} =  -(\tilde{a}_{\vartheta\vartheta}+
\tilde{b}_{\vartheta r\vartheta}/r)}&
\nonumber\\
&\displaystyle{\alpha_{\vartheta\varphi}  =  \alpha_{\varphi\vartheta}=
-1/2\,(\tilde{a}_{\vartheta\varphi}+\tilde{a}_{\varphi\vartheta}+
\tilde{b}_{\varphi r\vartheta})
\,, \quad
\alpha_{\varphi\varphi} =  -\tilde{a}_{\varphi\varphi}}&
\nonumber\\
&\displaystyle{\beta_{rr} = -1/2\,\tilde{b}_{r\varphi\vartheta}
\,, \quad
\beta_{r\vartheta} = \beta_{\vartheta r}=
1/4\,(\tilde{b}_{r\varphi r}-\tilde{b}_{\vartheta\varphi\vartheta})
\,, \quad
\beta_{r\varphi} = \beta_{\varphi r}=
1/4\,(\tilde{b}_{rr\vartheta}-\tilde{b}_{\varphi\varphi\vartheta}
-\tilde{b}_{r\vartheta r})}&
\nonumber\\
&\displaystyle{\beta_{\vartheta\vartheta} = 1/2\,\tilde{b}_{\vartheta\varphi r}
\,, \quad
\beta_{\vartheta\varphi} = \beta_{\varphi\vartheta}=
1/4\,(\tilde{b}_{\vartheta r\vartheta}+\tilde{b}_{\varphi\varphi r}-
\tilde{b}_{\vartheta\vartheta r})
\,, \quad
\beta_{\varphi\varphi} =
1/2\,(\tilde{b}_{\varphi r\vartheta}-\tilde{b}_{\varphi\vartheta r})}&
\nonumber\\
&\displaystyle{\gamma_r = 1/2\,(\tilde{a}_{\vartheta\varphi}-
\tilde{a}_{\varphi\vartheta}-\tilde{b}_{\varphi r\vartheta}/r)
\,, \quad
\gamma_\vartheta = 1/2\,(\tilde{a}_{\phi r}
-\tilde{b}_{\varphi\vartheta\vartheta}/r-\tilde{a}_{r\varphi})}&
\label{eq5:7}\\
&\displaystyle{\gamma_\varphi = 1/2\,(\tilde{a}_{r\vartheta}+\tilde{b}_{rr\vartheta}/r-
\tilde{a}_{\vartheta r}+\tilde{b}_{\vartheta\vartheta\vartheta}/r)}&
\nonumber\\
&\displaystyle{\delta_r = 1/4\,(\tilde{b}_{\vartheta\vartheta r}-\tilde{b}_{\vartheta
  r\vartheta}+\tilde{b}_{\varphi\varphi r})
\,, \quad
\delta_\vartheta = 1/4\,(\tilde{b}_{rr\vartheta}-\tilde{b}_{r\vartheta
  r}+\tilde{b}_{\varphi\varphi\vartheta})
\,, \quad
\delta_\varphi = -1/4\,(\tilde{b}_{r\varphi r}+
\tilde{b}_{\vartheta\varphi\vartheta})}&
\nonumber\\
&\displaystyle{\kappa_{\kappa rr} = -\tilde{b}_{\kappa rr}
\,, \quad
\kappa_{\kappa r\vartheta} = \kappa_{\kappa\vartheta r}=
-1/2\,(\tilde{b}_{\kappa r\vartheta}+\tilde{b}_{\kappa\vartheta r})
\,, \quad
\kappa_{\kappa r\varphi} = \kappa_{\kappa\varphi r} =
-1/2\,\tilde{b}_{\kappa\varphi r}}&
\nonumber\\
&\displaystyle{\kappa_{\kappa\vartheta\vartheta} =
-\tilde{b}_{\kappa\vartheta\vartheta}
\,, \quad
\kappa_{\kappa\vartheta\varphi} = \kappa_{\kappa\varphi\vartheta}=
-1/2\,\tilde{b}_{\kappa\varphi\vartheta}
\,, \quad
\kappa_{\kappa\varphi\varphi} = 0 \,.}&
\nonumber
\end{eqnarray}

\section{Determination of mean-field coefficients}
\label{sec:4}

For the determination of the mean-field coefficients $\tilde{a}_{\kappa\lambda}$,
$\tilde{b}_{\kappa\lambda r}$ and $\tilde{b}_{\kappa\lambda\vartheta}$ from the results of the
direct numerical simulations, two different approaches have been used, which are explained in the
following.

\subsection{A numerical approach (approach I)}
\label{sec:4:1}

We start from equation (\ref{eq3:8}) for $\mathbf{b}$ but interpret $\bm$ as a steady ``test field"
$\bmT$. More precisely, we require that $\mathbf{b}$ satisfies
\begin{equation}
  \frac{\partial\mathbf{b}}{\partial t}
  -\curl (\vm\times\mathbf{b})
  -\curl\mathbf{G}-
  \eta\nabla^2\mathbf{b}= \curl(\mathbf{v}\times\bmT) \,
  \label{eq4:2}
\end{equation}
inside the conducting shell and continues as a potential field in both parts of its surroundings.
The initial conditions loose their importance after some transient period.
%when the simulations reach a statistical equilibrium.
Calculating $\emf$ numerically according to (\ref{eq3:4}) and
(\ref{eq4:2}) for a given $\bmT$ and inserting the result in (\ref{eq3:12}) provides us with three
equations for the wanted 27 coefficients. We therefore carry out such calculations with the same
$\vm$ and $\mathbf{v}$ but nine different test fields $\bmT^{(i)}$ to obtain nine mean
electromotive forces $\emf^{(i)}$, $i = 1, 2, \dots, 9$. As far as the conditions for the validity
of (\ref{eq3:12}) are fulfilled we have then
\begin{equation}
{\mathcal{E}_\kappa}^{(i)} = \tilde{a}_{\kappa\lambda} {\overline{B}}^{(i)}_{\mathrm{T} \lambda}
    + \tilde{b}_{\kappa \lambda r} \frac{\partial {\overline{B}}^{(i)}_{\mathrm{T} \lambda}}{\partial r}
    + \tilde{b}_{\kappa \lambda \vartheta} \frac{1}{r}
    \frac{\partial {\overline{B}}^{(i)}_{\mathrm{T} \lambda}}{\partial \vartheta} \,,
    \quad i = 1, 2, \dots, 9 \, .
\label{eq4:4}
\end{equation}
The $\tilde{a}_{\kappa\lambda}$, $\tilde{b}_{\kappa\lambda r}$ and $\tilde{b}_{\kappa\lambda
\vartheta}$ depend on $\vm$ and $\mathbf{v}$ but not on $\bmT$, that is, they do not depend on $i$.
Clearly (\ref{eq4:4}) represents for each fixed $\kappa$ a set of nine linear equations for the
nine coefficients $\tilde{a}_{\kappa\lambda}$, $\tilde{b}_{\kappa\lambda r}$ and
$\tilde{b}_{\kappa\lambda \vartheta}$. The ${\mathcal{E}_\kappa}^{(i)}$ and
${\overline{B}}^{(i)}_{\mathrm{T} \lambda}$ occurring in these equations are then known quantities.
Provided the $\bmT^{(i)}$ are properly chosen, we may solve (\ref{eq4:4}) to find all
$\tilde{a}_{\kappa\lambda}$, $\tilde{b}_{\kappa\lambda r}$ and $\tilde{b}_{\kappa\lambda
\vartheta}$. The quantities $\emf$ and $\bmT$ as well as the $\tilde{a}_{\kappa\lambda}$,
$\tilde{b}_{\kappa\lambda r}$ and $\tilde{b}_{\kappa\lambda \vartheta}$ are in general functions of
position and time. The described procedure can be applied to all positions and all times.

In what follows $\vm$ and $\mathbf{v}$ will be extracted from the numerical simulations with
the models defined in section~\ref{sec:2}, that is, by the equations (\ref{eq2:2})-(\ref{eq2:10}).
Technically, parallel to the numerical solution of a problem as defined there, also the nine
solutions $\mathbf{b}$ of (\ref{eq4:2}) have been calculated.

There are some constraints on the choice of the test fields $\bmT^{(i)}$. Of course, they have to
be axisymmetric. They also have to be linearly independent. Otherwise there are no unique solutions
of the equations (\ref{eq4:4}). We further have to require that all higher than first-order spatial and
all time derivatives of the test fields are equal to zero, or at least sufficiently close to zero,
since otherwise (\ref{eq4:4}) is no longer justified. Fortunately, however, unlike $\bm$, the test
fields $\bmT^{(i)}$ need neither to be solenoidal nor to satisfy any boundary conditions. This
follows from the fact that relation (\ref{eq3:10}) can be derived using nothing else than the
definition $\emf = \vxb$ and an equation for $\mathbf{b}$ that formally agrees with
(\ref{eq3:8}), in which $\bm$, however, is understood as any axisymmetric vector field. A set of
test fields $\bmT^{(i)}$ used in our calculations is given in table \ref{tab4:2}. Note that not all
of these vector fields are regular at the polar axis. Consequently,
$\curl(\mathbf{v}\times\bmT)$ could become singular if the axis were included in the grid and
$\mathbf{v}$ were different from zero there. We also experimented with other test fields and
verified that the mean-field coefficients are independent of their particular choice as long as the
above constraints are obeyed.

\begin{table}
    \tbl{A set of test fields $\bmT^{(i)}$ used for the determination
    of $\tilde{a}_{\kappa\lambda}$ and $\tilde{b}_{\kappa\lambda r},
    \tilde{b}_{\kappa\lambda\vartheta}$.}
    {\begin{tabular}{cccccccccc}
        \toprule
        $i$ & $\enspace 1 \enspace$ & $\enspace 2 \enspace$ &
        $\enspace 3 \enspace$ & $\enspace 4 \enspace$ & $\enspace
        5 \enspace$ & $\enspace 6 \enspace$ & $\enspace 7 \enspace$
        & $\enspace 8 \enspace$ & $\enspace 9 \enspace$\\
        \colrule
        $\overline{B}_{\mathrm{T}r}^{(i)}$ & $1$ & $0$ & $0$ & $r$ & $0$ & $0$ & $\vartheta$ & $0$ & $0$ \\
        $\overline{B}_{\mathrm{T}\vartheta}^{(i)}$ & $0$ & $1$ & $0$ & $0$ & $r$ & $0$ & $0$ & $\vartheta$ & $0$ \\
        $\overline{B}_{\mathrm{T}\varphi}^{(i)}$ & $0$ & $0$ & $1$ & $0$ & $0$ & $r$ & $0$ & $0$ & $\vartheta$ \\
        \botrule
      \end{tabular}}
  \label{tab4:2}
\end{table}

As explained above, in the determination of the mean electromotive force $\emf$ often SOCA is used.
It is defined by cancelling the term with $\mathbf{G}$ in (\ref{eq4:2}). Our procedure for the
calculation of the $\tilde{a}_{\kappa\lambda}$, $\tilde{b}_{\kappa\lambda r}$ and
$\tilde{b}_{\kappa\lambda \vartheta}$ also works on this level.

\subsection{A semi-analytical approach using SOCA (approach II)}
\label{sec:4:2}

We start again from equation (\ref{eq3:8}) but introduce some simplifications so that the remaining
equation for $\mathbf{b}$ allows an analytical solution. In that sense we restrict ourselves to the
case $\vm = \mathbf{0}$. Furthermore we accept the second-order correlation approximation and
cancel the term with $\mathbf{G}$. Finally we consider only the steady case, that is, assume
$\mathbf{v}$, $\mathbf{b}$ and also $\bm$ to be independent of time. With these assumptions
equation (\ref{eq3:8}) turns into
\begin{equation}
\eta\nabla^2\mathbf{b}=-\bnabla\times(\mathbf{v}\times\bm).
\label{eq4:10}
\end{equation}

In the solutions of the problems defined in section~\ref{sec:2} the velocity $\mathbf{v}$ is
represented in the form
\begin{eqnarray}
\mathbf{v} &=& - \bnabla \times (\mathbf{r} \times \bnabla \phi) - \mathbf{r} \times \bnabla \psi
\label{eq4:12}
\end{eqnarray}
with scalars $\phi$ and $\psi$ given by
\begin{equation}
\phi = \sum_{l,m} \, \phi_l^m (r) Y_l^m (\vartheta, \varphi) \, , \quad
    \psi = \sum_{l,m} \,  \psi_l^m (r) Y_l^m (\vartheta, \varphi) \, .
\label{eq4:14}
\end{equation}
The $\phi_l^m$ and $\psi_l^m$ are complex functions of $r$ satisfying $\phi_l^{m *} = \phi_l^{-m}$
and $\psi_l^{m *} = \psi_l^{-m}$, but $\phi_l^0 = \psi_l^0 = 0$. The  $Y_l^m$ are spherical
harmonics, $Y_l^m (\vartheta, \varphi) = P_l^{|m|} (\cos \vartheta) \exp(\mbox{i} m \varphi)$, with
$P_l^m$ being associated Legendre polynomials. In the following the $\phi_l^m$ and $\psi_l^m$ are
considered as given.

In appendix~\ref{app:1} the solution of equation (\ref{eq4:10}) for $\mathbf{b}$ is derived, that
is, $\mathbf{b}$ is expressed by the $\phi_l^m$, $\psi_l^m$, $Y_l^m$ and the components of
$\bm$. On this basis $\emf$ has been calculated. It occurs at first in a form analogous to
(\ref{eq3:10}), more precisely
\begin{equation}
\mathcal{E}_\kappa (r, \vartheta) = \int_\mathcal{V} K_{\kappa \lambda} (r, \vartheta; r', \vartheta')
    {\overline{B}}_{\lambda} (r', \vartheta') \, \mbox{d} v' \, .
\label{eq4:50}
\end{equation}
The kernel $K_{\kappa \lambda}$ is determined by the $\phi_l^m (r)$, $\psi_l^m (r)$ and the $P_l^m
(\cos \vartheta)$. As an example, $K_{rr}$ is explicitly given in appendix~\ref{app:1}.

Knowing the kernel $K_{\kappa \lambda}$ we may calculate the ${\tilde{a}}_{\kappa \lambda}$
according to
\begin{equation}
{\tilde{a}}_{\kappa \lambda} = \int_\mathcal{V} K_{\kappa \lambda} (r, \vartheta; r', \vartheta')
\, \mbox{d} v' \, . \label{eq4:51}
\end{equation}
This relation turns into those for ${\tilde{b}}_{\kappa \lambda r}$ or ${\tilde{b}}_{\kappa \lambda
\vartheta}$ if the factors $(r' - r)$ or $r (\vartheta' - \vartheta)$, respectively, are
additionally inserted into the integrand. Note that $\bm$ does not enter the expression for
$K_{\kappa\lambda}$ and that the mean-field coefficients are thus independent of the mean field.

In fact only the ${\tilde{a}}_{\kappa \lambda}$ have been calculated so far. It was found, e.g.,
\begin{eqnarray}
\tilde{a}_{rr}(r,\vartheta) &=& \frac{2}{\eta} \sum_{l,l'\,;\,m>0} \Bigg\{
\int_{r_i}^{r_0}\bigg[{\hat{f}}_l(r,r')\,\mbox{Re}\Big(\psi^{m*}_{l'}(r)
   {\hat{\phi}}^m_{l}(r')\Big) \nonumber\\
&& +\tilde{g}_l(r,r')\,
   \mbox{Re}\Big({\hat{\phi}}^{m*}_{l'}(r)\psi^m_{l}(r')\Big)\bigg] {r'}^2
   \mbox{d}r' \; R^m_{l'l}(\vartheta) \nonumber\\
&& -\int_{r_i}^{r_0}\bigg[ {\hat{f}}_l(r,r')\,
   \mbox{Im}\Big(\hat{\phi}^{m*}_{l'}(r)\hat{\phi}^m_l(r')\Big)
   -\tilde{g}_l(r,r')\,\mbox{Im}\Big(\psi^{m*}_{l'}(r)\psi^m_l(r')\Big)
   \bigg] \,{r'}^2\mbox{d}r' \nonumber\\
&& \times m\Big(Q^m_{l'l}(\vartheta)+Q^m_{ll'}(\vartheta)\Big)\big/
   \sin\vartheta \Bigg\} \, ,
\label{eq4:58}
\end{eqnarray}
where ${\hat{f}}_l(r,r')$ and ${\tilde{g}}_l(r,r')$ are Green's functions, and
$R^m_{l'l}(\vartheta)$ and $Q^m_{l'l}(\vartheta)$ specific combinations of associated Legendre
polynomials $P_l^m (\cos \vartheta)$, all explained in appendix~\ref{app:1}. The other
${\tilde{a}}_{\kappa \lambda}$ are also given there.

\section{Mean-field coefficients for rotating magnetoconvection and a geodynamo model}
\label{sec:5}

\subsection{Rotating magnetoconvection}
\label{sec:5:1}

\begin{figure}[t]
    \centering\includegraphics[scale=0.8,angle=90]{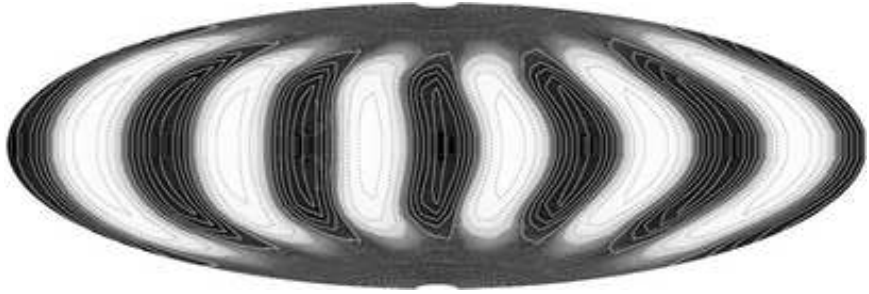}
    \caption{The radial velocity in the magnetoconvection case with 6-fold azimuthal symmetry
    at $r=0.59\,r_0$, normalized with its maximum given by $V_r=16.98\,\eta/D$. In the grey
    scale coding, white and black correspond to $-1$ and $+1$, that is, downflows and upflows,
    respectively, and the contour lines to $\pm 0.1,\,\pm 0.3,\,\pm 0.5,\,\pm 0.7,\, \pm 0.9$.}
    \label{fig:u:magnetoconvection}
\end{figure}

The first example considered is adopted from \cite{olson_99} with the governing parameters
$\Ek=3\times 10^{-4}$, $\Ra=94(= 1.5 \Ra_c$, where $\Ra_c$ means the critical value of $\Ra$) and
$\Pr=\Pm=1$. Moreover, an axisymmetric toroidal magnetic field is imposed via an inhomogeneous
boundary condition of the form
\begin{equation}
  B_\varphi=B_0\sin(2\vartheta)\quad\mbox{at}\enspace
  r=r_i,\,\,\,\,\,r=r_0 \,.
  \label{eq6:2}
\end{equation}
Figure \ref{fig:u:magnetoconvection} shows the radial velocity field at $r=0.59\,r_0$ for an
Elsasser number of the imposed field $\Lambda_0=1$, where $\Lambda_0$ is defined according to (6)
but with $B$ replaced by $B_0$. A typical columnar convection pattern is revealed. Apart from a
steady azimuthal drift of the convection columns the flow and also the magnetic field are steady.
In addition, the azimuthal drift and so the remaining time dependence can be removed by a
transformation to a corotating frame of reference. The electromotive force $\emf$ is identical in
both the original and the corotating frame \citep[e.g.,][]{raedleretal06}.

\begin{figure}[t]
     \centering\includegraphics[scale=1.0]{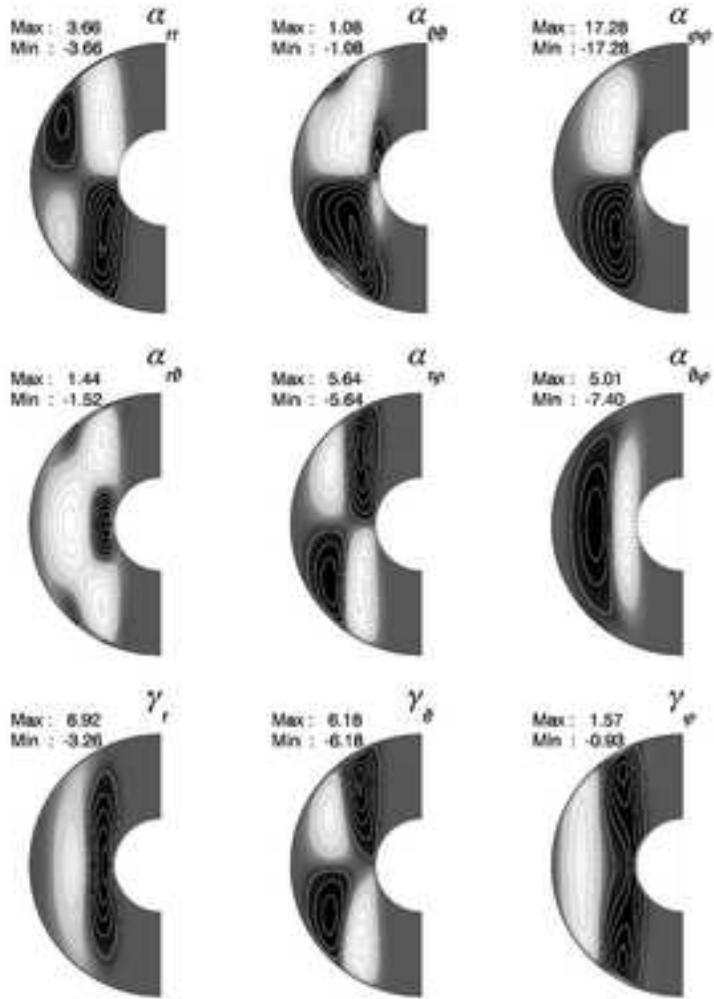}
     \caption{Components of the symmetric $\boldsymbol{\alpha}$-tensor and the
     $\boldsymbol{\gamma}$-vector  in a meridional plane in the magnetoconvection example,
     determined by method I, in units of $\eta/D$. For each component the grey scale (white --
     negative, black -- positive values) is separately adjusted to its maximum modulus.
     Note the negative signs in the definitions of $\boldsymbol{\alpha}$ and $\boldsymbol{\gamma}$
     in equation (\ref{eq5:4}).}
    \label{fig:alpha:magnetoconvection}
\end{figure}

The magnetic Reynolds number $\Rm$ is about 12, that is too low for the onset of self-sustaining
dynamo action. Nevertheless, fundamental effects of the convection, which are of high interest for
a dynamo, can be analysed in terms of mean fields, for example the generation of a poloidal from a
toroidal field. The ratio of magnetic to kinetic energy density is about 10, and the resulting
field strength is described by $\Lambda\approx0.6$.

\begin{figure}[t]
    \centering\includegraphics[scale=1.0]{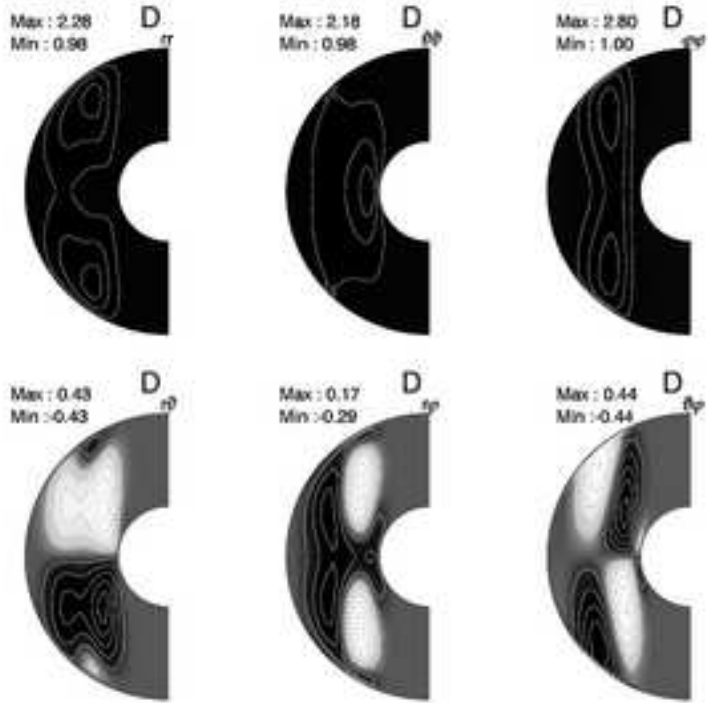}
    \caption{The symmetric diffusivity tensor $\mathbf{D}$ in the meridional plane in the
    magnetoconvection example, determined by method I, in units of $\eta$. Grey scales as in
    figure 2.}
    \label{fig:diffusion}
\end{figure}

Figure \ref{fig:alpha:magnetoconvection} shows the six independent components of the tensor
$\boldsymbol{\alpha}$ and the three components of the vector $\boldsymbol{\gamma}$ in a
meridional plane derived by approach I.
%%They do not completely coincide with the corresponding
%%results of approach II. This is not surprising since approach II is based on SOCA. In the
%%steady case considered here, it can only be justified for $\Rm\ll 1$. The two results come into
%%agreement if $\Rm$ is scaled down. Alternatively, approach I can be applied cancelling the term
%%$\nabla\times G$ and $\vm$ in (\ref{eq4:2}). Then, the resulting mean-field coefficients given
%%by approach I are again in good agreement with those derived by means of approach II.
All mean-field coefficients are essentially determined by the columnar convection outside the inner
core tangent cylinder. As a consequence of the symmetry properties of the velocity field and the
induction equation, all mean-field coefficients are either symmetric or antisymmetric with respect
to the equatorial plane. The diagonal components of $\boldsymbol{\alpha}$, for instance, are
antisymmetric, in its major contributions negative in the northern and positive in the southern
hemisphere. Among the $\boldsymbol{\alpha}$-components, $\alpha_{\varphi\varphi}$ dominates,
indicating that the generation of a poloidal field from a toroidal one is more effective than the
reversed process. However, due to the other non-vanishing components, especially $\alpha_{rr}$,
$\alpha_{r\vartheta}$ and $\alpha_{\vartheta\vartheta}$, generation of toroidal field by an
$\alpha$-effect also takes place.

The mean-field diffusivity tensor $\mathbf{D}$ is given by
\begin{equation}
D_{\kappa\lambda}=\eta\delta_{\kappa\lambda}+\beta_{\kappa\lambda} \, .
\label{eq6:20}
\end{equation}
Its components are shown in figure~\ref{fig:diffusion}. Although the molecular magnetic diffusivity
is rather high ($\Pm=1$) and the vigour of the convection is rather low ($\Ra=1.5\Ra_c$), the
turbulent diffusion is of the same order as the molecular one in the convection region.

The diffusivity tensor $\mathbf{D}$ has the interesting property of being positive definite
everywhere. This can be concluded from the fact that all its diagonal elements as well as all
sub-determinants are positive. As explained in appendix~\ref{app:2}, this property implies that the
induction effects expressed by $\mathbf{D}$ do not contribute to a growth of the total
magnetic energy stored in the mean magnetic field but favour its dissipation. Because the fact that
$\beta_{\kappa\lambda}$ has been defined with some arbitrariness, this statement has, however, to
be considered with some caution (see also appendix~\ref{app:2}).

The $\boldsymbol{\delta}$-vector and the $\boldsymbol{\kappa}$-tensor (which are not displayed
here) have been derived as well, and they are used in the magnetoconvection and dynamo calculations
of section~\ref{sec:6}.

\begin{figure}[t]
   \centering\includegraphics[scale=0.4]{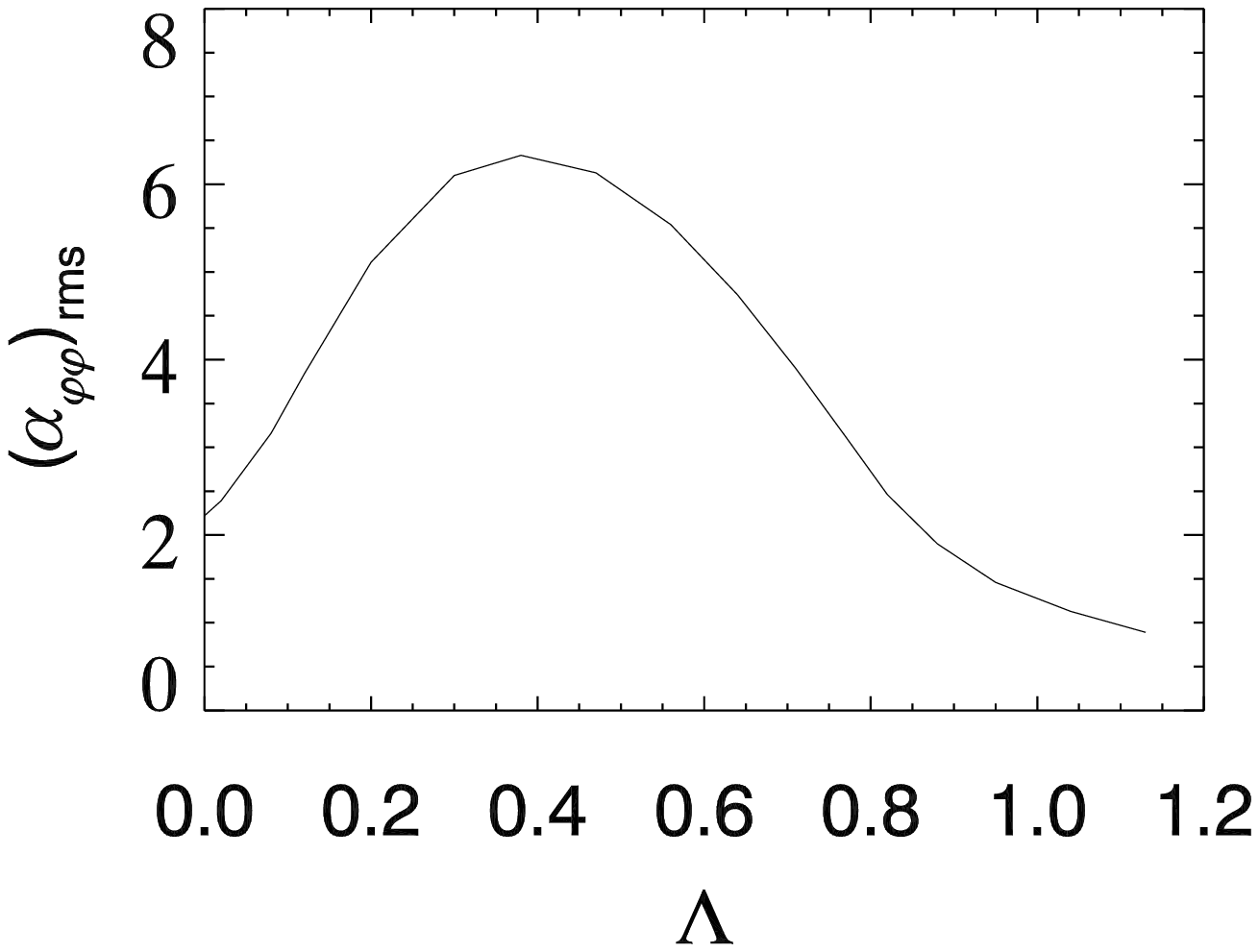}\hspace{2mm}\includegraphics[scale=0.4]{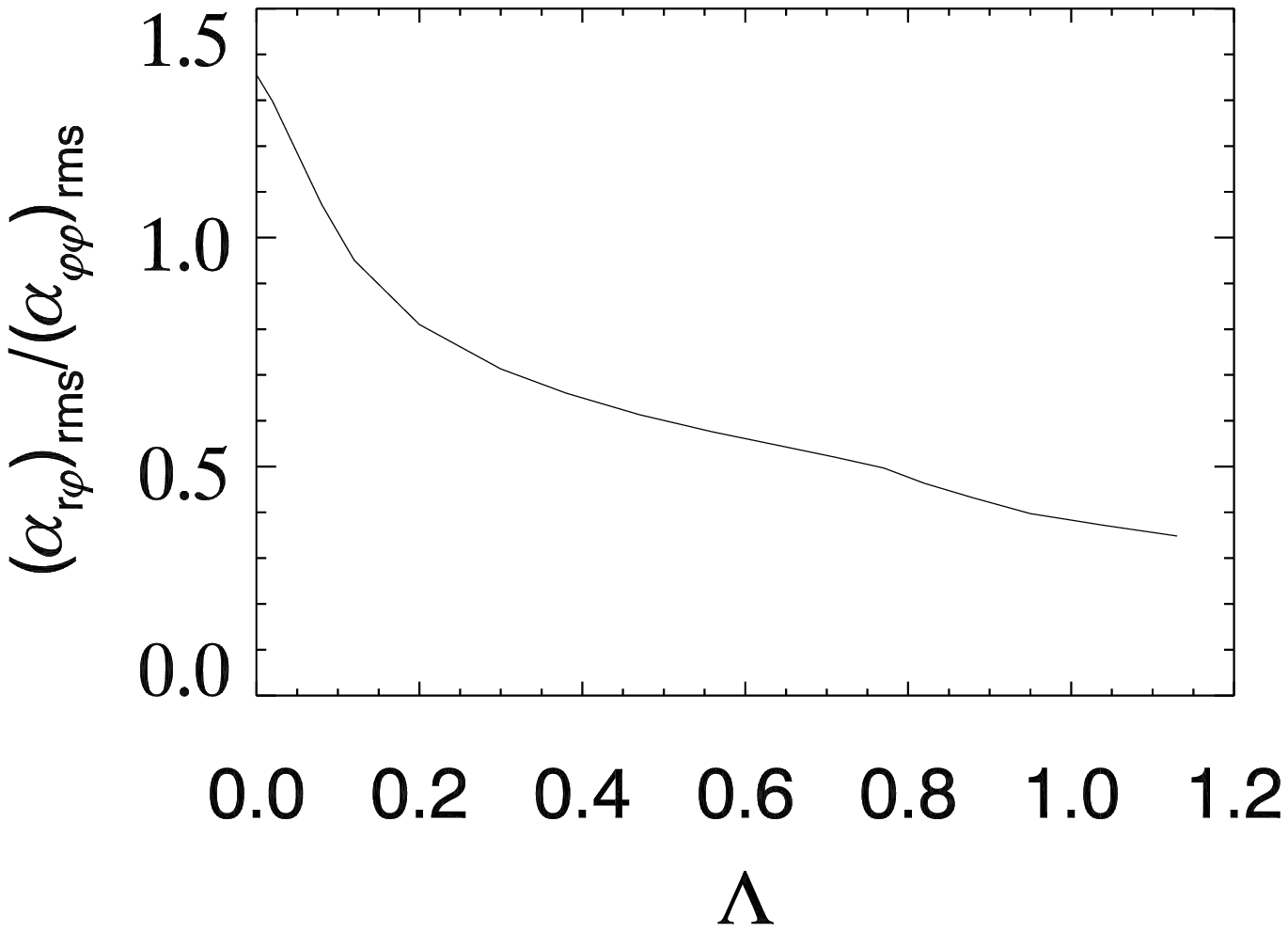}\hspace{2mm}\includegraphics[scale=0.4]{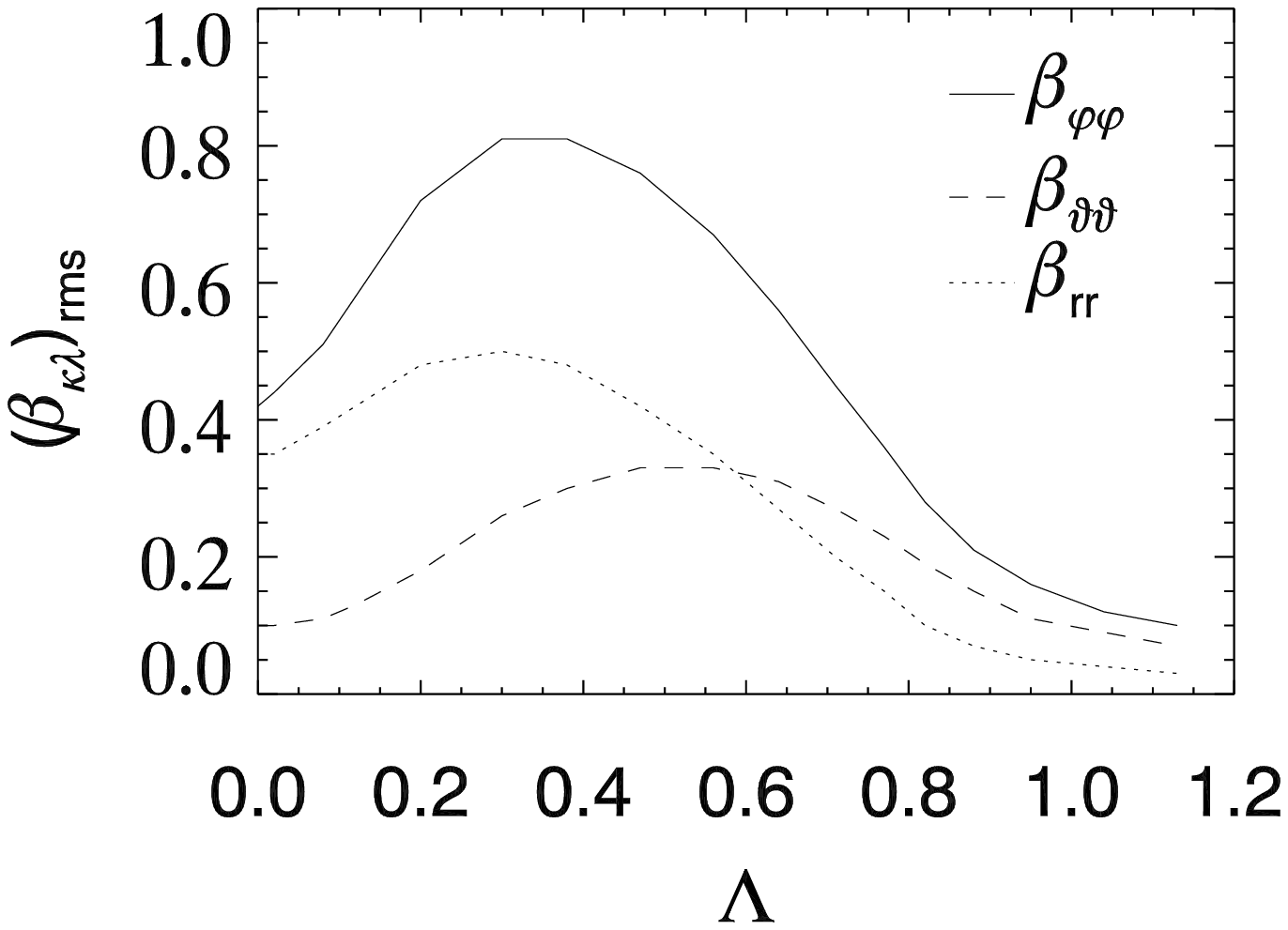}
   \caption{$(\alpha_{\varphi\varphi})_\mathrm{rms}$ in units of $\eta/D$ (left), the ratio
   $(\alpha_{r\varphi})_\mathrm{rms}/(\alpha_{\varphi\varphi})_\mathrm{rms}$ (middle), and the
   diagonal components of $(\beta_{\kappa\lambda})_\mathrm{rms}$ in units of $\eta$ (right) in the
   magnetoconvection case, in dependence on the Elsasser number $\Lambda$.}
   \label{fig:quenching}
\end{figure}

\subsection{Quenching of mean-field coefficients}
\label{sec:5:2}

We point out that the velocity $\mathbf{v}$ needed for the determination of the mean-field
coefficients $\tilde a_{\kappa\lambda}$, $\tilde b_{\kappa\lambda r}$ and $\tilde
b_{\kappa\lambda\vartheta}$ were taken from simulations with non-zero mean magnetic field $\bm$.
Therefore, the resulting coefficients are already subject to a magnetic quenching corresponding to
this magnetic field. In this respect, results obtained with various $\bm$, measured by the Elsasser
number $\Lambda$, are of interest. Figure \ref{fig:quenching} (left) shows
$(\alpha_{\varphi\varphi})_\mathrm{rms}$, the r.m.s. value of $\alpha_{\varphi\varphi}$, as a
function of $\Lambda$. The increase of this quantity in the presence of a weak magnetic field, that
is for small $\Lambda$, is due to an increasing vigor of convection by the relaxation of the
geostrophic constraint \citep{fearn_98}. A strong magnetic field however inhibits convection and
reduces $(\alpha_{\varphi\varphi})_\mathrm{rms}$. The kinetic energy of the convection varies with
$\Lambda$ similar to $(\alpha_{\varphi\varphi})_\mathrm{rms}$. The other $\boldsymbol{\alpha}$
components are also quenched. The quenching is however not the same for different components,
leading to varying amplitude relations among these components for varying strength of the mean
magnetic field (figure~\ref{fig:quenching} middle). In addition to the
$\boldsymbol{\alpha}$-quenching, e.g., also a $\boldsymbol{\beta}$-quenching takes place (figure
\ref{fig:quenching} right).

\begin{figure}[t]
    \centering\includegraphics[scale=1.0]{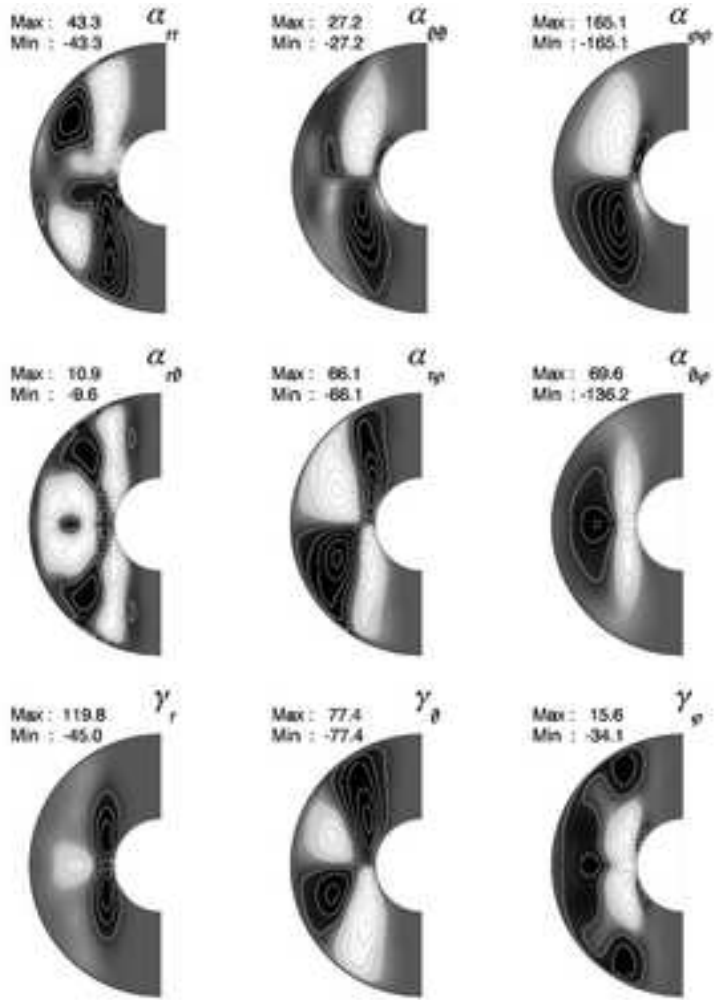}
    \caption{Components of the symmetric $\boldsymbol{\alpha}$-tensor and the
    $\boldsymbol{\gamma}$-vector in a meridional plane in the geodynamo example,
    determined by method I, in units of $\eta/D$. Grey scales as in figure 2.}
    \label{fig:alpha:geodynamo}
\end{figure}

\subsection{A quasi-steady geodynamo}
\label{sec:5:3}

As a further example a quasi-steady geodynamo model is examined, which has been used before as a
numerical dynamo benchmark \citep{christensen_01}. The governing parameters have been chosen to be
$\Ek=10^{-3}$, $\Ra=100\,(=1.79\Ra_c)$, $\Pr=1$ and $\Pm=5$. The columnar convection
pattern is similar to that in the magnetoconvection example (figure \ref{fig:u:magnetoconvection}),
but with a natural 4-fold azimuthal symmetry. The intensity of the fluid motion is characterised by
$\Rm\approx 40$, and the magnetic energy density exceeds the kinetic one by a factor of 20.
Again, except for an azimuthal drift of the convection columns, the velocity field is stationary.

The components of the $\boldsymbol{\alpha}$-tensor and the $\boldsymbol{\gamma}$-vector are
shown in figure \ref{fig:alpha:geodynamo}. Among the $\boldsymbol{\alpha}$-components,
$\alpha_{\varphi\varphi}$ again dominates, indicating a very efficient generation of poloidal
from toroidal magnetic field. The components $\alpha_{rr}$, $\alpha_{r\vartheta}$ and
$\alpha_{\vartheta\vartheta}$ are somewhat lower in amplitude. Since the influence of the
differential rotation on the generation of toroidal field is negligible, this example can be
classified as an $\alpha^2$-dynamo. The imbalance in the amplitudes of the $\alpha$-components is
reflected in the larger strength of the mean poloidal field compared to the mean toroidal field. As
before in the example of magnetoconvection, the $\boldsymbol{\gamma}$-effect acts to expel flux
from the central dynamo region where the convection takes place.

\begin{figure}[t]
    \centering\includegraphics[scale=1.0]{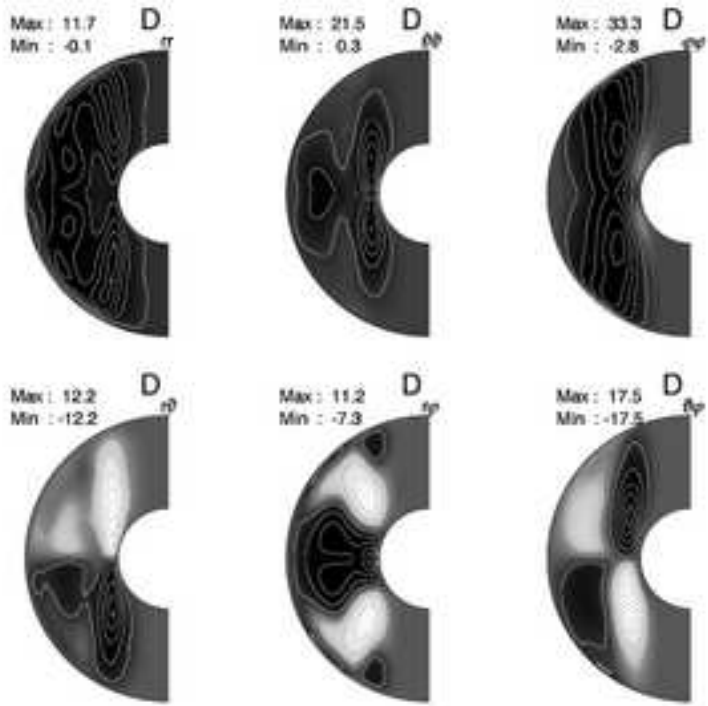}
    \caption{Components of the symmetric diffusivity tensor $\mathbf{D}$ in a meridional plane
    in the geodynamo example, determined by method I, in units of $\eta$. Grey scales as
    in figure 2.}
    \label{fig:diffusion:geodynamo}
\end{figure}

\begin{figure}[t]
    \centering\includegraphics[scale=1.0]{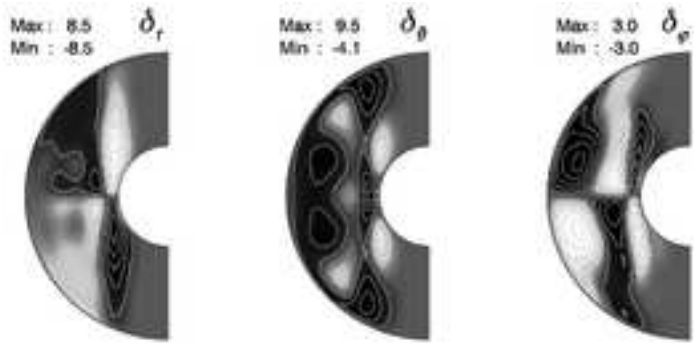}
    \caption{Components of the $\boldsymbol{\delta}$-vector in a meridional plane
    in the geodynamo example, determined by method I, in units of $\eta$. Grey scales as in
    figure 2.}
    \label{fig:delta:geodynamo}
\end{figure}

The corresponding diffusivity tensor $\mathbf{D}$ is displayed in figure
\ref{fig:diffusion:geodynamo}. The turbulent diffusivity $\boldsymbol{\beta}$ exceeds the molecular
one, $\eta$, by more than a factor of 10 in the convection region outside the inner core tangent
cylinder, leading to a very efficient diffusion of the mean magnetic field.

There is a weak negative contribution to $D_{\varphi\varphi}$ at the inner boundary close to the
equator, which is negligible in the mean-field model of section~\ref{sec:6:3}. The diffusivity
tensor $\mathbf{D}$ thus slightly deviates from being positive definite. We argue in appendix
\ref{app:2} that another than our particular choice $\slb_{\kappa\lambda\varphi}=0$ will lead to
another $\mathbf{D}$ without changing $\emf$ and thus the physical situation. Furthermore, a
parametrisation of $\emf$ considering higher than first-order derivatives of $\bm$ will also lead
to changes of the low-order mean-field coefficients. In section~\ref{sec:6:4} we will see the need
of a better parametrisation of $\emf$ in the geodynamo case.

As for the $\delta$-effect we recall that the combination of this effect with a mean rotational
shear may constitute a dynamo \citep{raedler_69b,raedler_70,raedler_86,roberts72,raedleretal03}. A
dynamo mechanism of that kind, however, can not play a dominant role here, since neither $\vm$ nor
$\boldsymbol{\gamma}$ imply a sufficiently strong shear. As we will see in section~\ref{sec:6:5}
below, however, the $\boldsymbol{\delta}$ terms, which are displayed in figure
\ref{fig:delta:geodynamo}, may diminish the decay of a mean magnetic field.

\subsection{Limitation of SOCA}
\label{sec:5:4}

The mean-field coefficients determined by approach I and approach II (SOCA) show for all $\Rm$ an
almost perfect congruence of their profiles. However, mean-field coefficients determined by means
of SOCA exhibit typically overestimated amplitudes for $\Rm\gtrsim 10$.

\begin{figure}[t]
    \centering\includegraphics[scale=0.75]{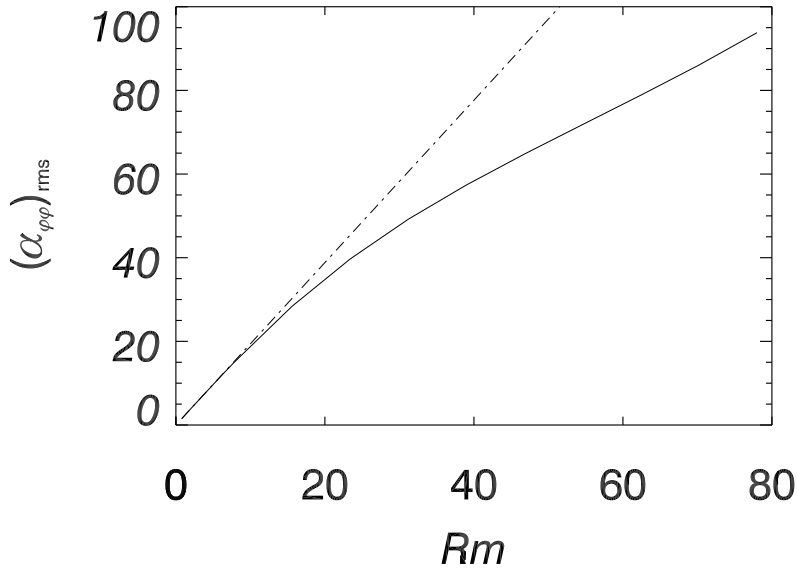}
    \caption{The quantity $(\alpha_{\varphi\varphi})_\mathrm{rms}$
    in the geodynamo case, in units of $\eta/D$, with
    $\alpha_{\varphi\varphi}$ determined by methods I and II
    (solid and dashed line, respectively) in dependence on $\Rm$.}
    \label{fig:soca}
\end{figure}

In figure \ref{fig:soca}, $(\alpha_{\varphi\varphi})_\mathrm{rms}$ is plotted versus $\Rm$. In all
calculations from which the mean-field coefficients were derived, $v$ was the same, and the
variation of $\Rm$ is only due to a variation of $\eta$ or $\Pm$ which, for the calculation of
$\mathbf{b}$ in equations (\ref{eq4:2}) or (\ref{eq4:10}), can be chosen differently from their
values in the original model simulations. For small $\Rm$ the results of approaches I and II
coincide and vary linearly with $\Rm$. For $\Rm \gtrsim 10$, the slope of
$(\alpha_{\varphi\varphi})_\mathrm{rms}$ determined by approach I flattens. In particular,
$\alpha_{\varphi\varphi}$ derived by approach I, that is, without restriction to SOCA, leads to
amplitudes which are, e.g., for $\Rm = 40$ about $30\%$ smaller than those gained by approach II,
that is, with SOCA calculations. The consequences for the dynamo action in a mean-field model are
studied in the following section.

Let us add some explanation for the linear dependence of $(\alpha_{\varphi\varphi})_\mathrm{rms}$
on $\Rm$ observed in both approaches for small $\Rm$, and in approach II for all $\Rm$. In SOCA,
when assuming a steady velocity $\mathbf{v}$, an $\boldsymbol{\alpha}$-component like
$\alpha_{\varphi\varphi}$, say simply $\alpha$, is given by $\alpha = f v^2 D / \eta$, where $f$ is
a purely numerical factor. We may write this also in the forms $\alpha = (f \eta / D) \Rm^2$ or
$\alpha = f v \Rm$. If $\eta$ and $D$ are fixed, $\alpha$ appears to be proportional to $\Rm^2$,
which may then vary with $v$. If, however, $v$ is fixed, $\alpha$ proves to be proportional to
$\Rm$, which may vary with $\eta$. This corresponds to the results presented in figure
\ref{fig:soca}. The deviation of the results for $(\alpha_{\varphi\varphi})_\mathrm{rms}$ obtained
in approach I from the linearity in $\Rm$ indicates that we are no longer in the range of
applicability of SOCA or that the time variation of $\mathbf{v}$ is no longer sufficiently weak.

The usually given sufficient condition for the applicability of SOCA in the limit of steady motions
reads $\Rm\ll 1$, where $\Rm$ is defined with a typical length of the fluid flow. Even if this
length is slightly overestimated by $D$ used in our definition of $\Rm$, our finding of the
applicability of SOCA for $\Rm\lesssim 10$ is very remarkable.

\section{Comparison between numerical simulations and mean-field models}
\label{sec:6}

\subsection{Mean-field model}
\label{sec:6:1}

In order to compare direct numerical simulations and mean-field theory, an axisymmetric mean-field
dynamo model involving all 27 mean-field coefficients $\tilde{a}_{\kappa\lambda}$,
$\tilde{b}_{\kappa\lambda r}$, and $\tilde{b}_{\kappa\lambda\vartheta}$ has been constructed.
The model also enables isolating certain dynamo processes.

Decomposing the axisymmetric mean magnetic field $\bm$ in its poloidal and toroidal parts,
\begin{equation}
\bm=\bm_\mathrm{pol}+\bm_\mathrm{tor}
\label{eq7:2}
\end{equation}
with
\begin{equation}
\bm_\mathrm{pol}=\bnabla\times A\mathbf{e}_\varphi\,,\quad
\bm_\mathrm{tor}= B\mathbf{e}_\varphi,
\label{eq7:4}
\end{equation}
where $\mathbf{e}_\varphi$ is the unit vector in azimuthal direction, we may write the
induction equation (\ref{eq3:2}) as
\begin{eqnarray}
\frac{\partial A}{\partial t} &=& \frac{1}{r\sin\vartheta}\overline{\mathbf{V}}_\mathrm{pol}
\cdot\bnabla(r\sin\vartheta A)+\mathcal{E}_\varphi-\eta\Delta' A \label{eq7:6}\\
\frac{\partial B}{\partial t} &=& r\sin\vartheta\,\overline{\mathbf{V}}_\mathrm{pol}
\cdot\bnabla\left(\frac{B}{r\sin\vartheta}\right)+\frac{1}{r}\frac{\partial({\overline{V}}_\varphi/r\sin
\vartheta,\,r\sin\vartheta A)}{\partial(r,\vartheta)}+(\bnabla\times\emf_\mathrm{pol})_\varphi
-\eta\Delta' B \,, \label{eq7:8}
\end{eqnarray}
where $\Delta'=\Delta-1/(r\sin\vartheta)^2$. Here, the notations
$\emf_\mathrm{pol}=(\mathcal{E}_r,\mathcal{E}_\vartheta,0)$ and
$\overline{\bv}_\mathrm{pol}=(\overline{V}_r,\overline{V}_\vartheta,0)$ have been used. $\emf$
in its dependence on $\bm$ has to be taken from (\ref{eq3:12}). The above equations are then solved
in a spherical shell with electrically insulating inner and outer surroundings. Thus, the mean
magnetic field is assumed to continue as a potential field in both parts exterior to the fluid
shell.

The two coupled equations (\ref{eq7:6})-(\ref{eq7:8}) are solved by a finite difference method on an
equidistant grid in radial and latitudinal direction. An alternating direction implicit scheme for
parabolic equations with mixed derivatives according to \cite{mckee} has been used to discretise
the equations. This enables an efficient implicit treatment of advection and diffusion terms,
whereas mixed and higher order derivatives are treated explicitly.

\subsection{Magnetoconvection}
\label{sec:6:2}

\begin{figure}[t]
    \centering\includegraphics[scale=0.9]{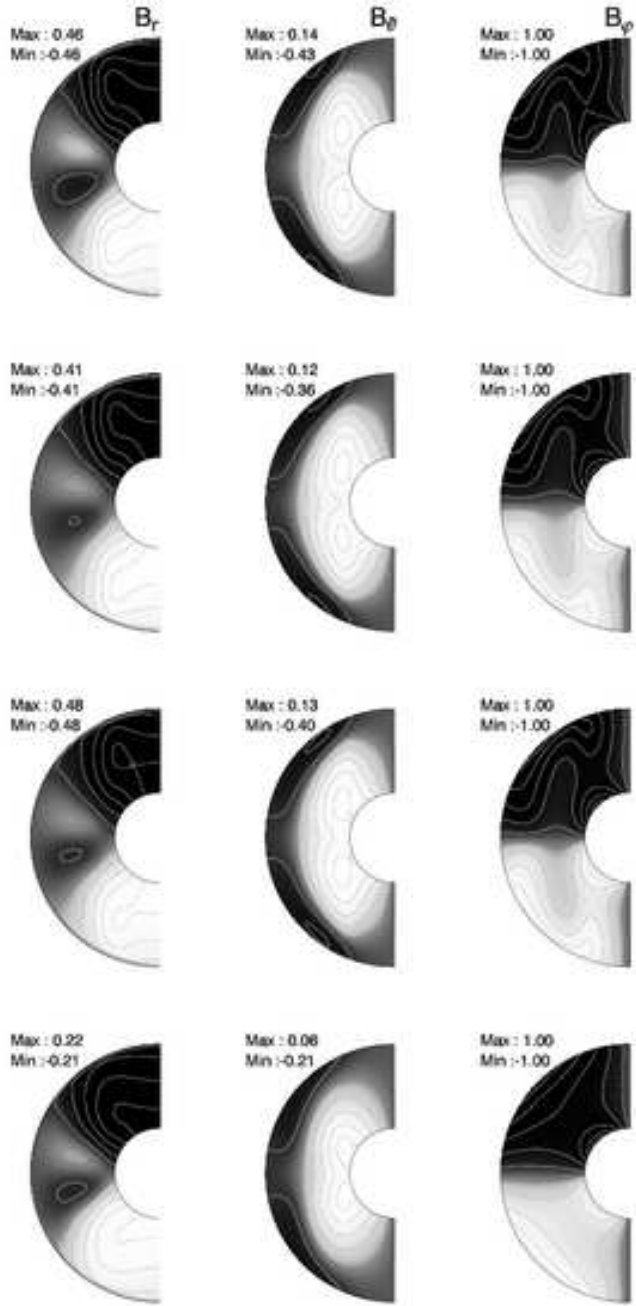}
    \caption{Comparison between numerical simulations and
      mean-field calculations in the example of magnetoconvection:
      azimuthally averaged magnetic field
      components resulting from a direct numerical
      simulation (first row), results given by the mean-field model
      based on all 27 coefficients derived by approach I (second
      row), mean-field calculation with coefficients
      derived applying SOCA (approach I with $\mathbf{G}=\mathbf{0}$, third row),
      and mean-field calculation with coefficients for isotropic turbulence
      (last row). Maxima and minima of the field components are given in
      units of $(\varrho\mu\eta\Omega)^{1/2}$. Grey scales as in figure 2.}
    \label{fig:magnetoconvection}
\end{figure}

How well do the results given by mean-field models match with the corresponding azimuthally
averaged fields determined by direct numerical simulations? Let us first consider the rotating
magnetoconvection model discussed in section~\ref{sec:5:1}. Figure \ref{fig:magnetoconvection}
presents a comparison between direct numerical simulations and mean-field calculations. In the
first row, the azimuthally averaged magnetic field components resulting from the direct numerical
simulation are shown. They correspond in great detail to the results of our mean-field model
(second row), in which all 27 mean-field coefficients have been used. The poloidal field is dipolar
with inverse flux spots near the equatorial plane, and the applied azimuthal field is expelled from
the region occupied by the convection columns.

A mean-field simulation relying on mean-field coefficients derived in SOCA (third row in figure
\ref{fig:magnetoconvection}) fits equally well. This reflects that mean-field coefficients given by
SOCA, even overestimated by a few per cent in their amplitudes though, still lead to a reliable
parametrisation of the mean electromotive force in this parameter regime. Moreover, amplitude
deviations simultaneous in $\boldsymbol{\alpha}$ and $\boldsymbol{\beta}$ might not strongly
influence the efficiency of the generation of poloidal from toroidal magnetic field and vice versa,
as suggested by a simple scaling analysis: The efficiency of these processes can be expressed by
the dimensionless number $P=\alpha_0^2\,D^2/\beta_0^2$. Here, $\alpha_0$ and $\beta_0$ mean
typical values for the $\alpha$-effect and the turbulent diffusivity, respectively, and $D$
stands for a typical length scale. Since $\boldsymbol{\alpha}$ and $\boldsymbol{\beta}$ are
likewise overestimated in their amplitudes, this factor cancels out and has no influence on $P$. As
a consequence, the resulting field resembles the mean field displayed in the second row, even
though the applied mean-field coefficients have larger amplitudes.

\begin{figure}[t]
    \centering\includegraphics[scale=0.9]{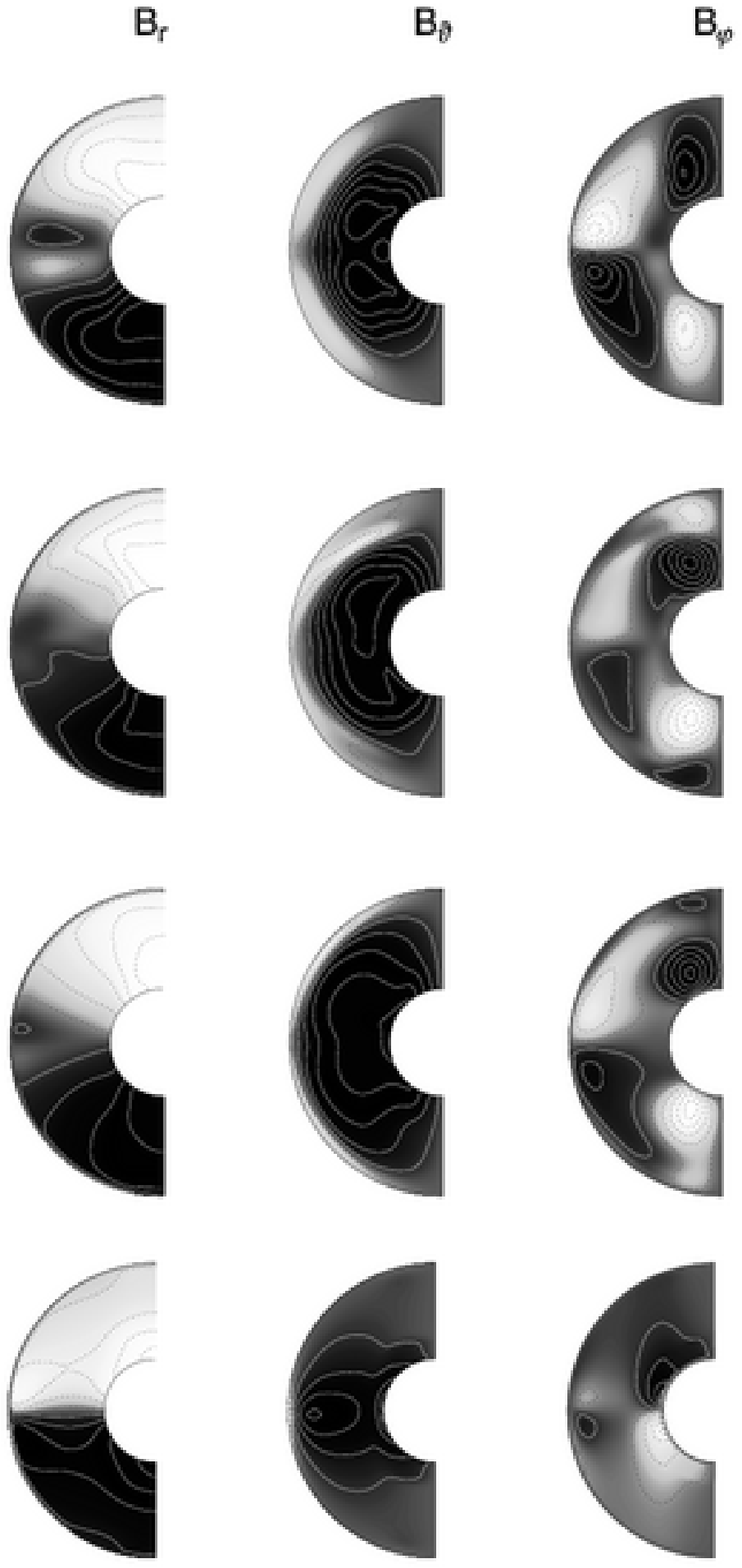}
    \caption{Comparison between numerical simulations and mean-field
    calculations in the example of the geodynamo: azimuthally
    averaged magnetic field components resulting from a direct numerical
    simulation (first row), results as given by mean-field modelling with coefficients derived by
    approach I (second row), mean-field calculation with coefficients in
    SOCA (approach I with $\mathbf{G}=\mathbf{0}$, third row), and with
    coefficients for isotropic turbulence (last row).
    Different from the solution of the direct numerical calculation which is
    stationary, all mean-field solutions decay exponentially with
    decay rates $\lambda=3.5\,\eta/D^2$ (second row), $\lambda=6.5\,\eta/D^2$
    (third row), and $\lambda\approx 75\,\eta/D^2$ (last row).}
    \label{fig:geodynamo}
\end{figure}

The mean field components shown in the last row of figure \ref{fig:magnetoconvection} have been
determined applying the isotropic approximation
$\alpha_{\lambda\kappa}=\alpha_I\delta_{\lambda\kappa}$,
$\beta_{\lambda\kappa}=\beta_I\delta_{\lambda\kappa}$. Concerning the coefficients $\alpha_I$ and
$\beta_I$ we rely on SOCA results for isotropic turbulence in the limit of steady motion (see,
e.g., \citet{raedler_80} or \citet{raedler00c}) and put
\begin{equation}
   \alpha_I= -\frac{1}{3\eta}\overline{\mathbf{a}\cdot(\curl\mathbf{a})}
   \quad\mathrm{and}\quad \beta_I=\frac{1}{3\eta}\overline{\mathbf{a}^2} \, ,
   \label{eq8:2}
\end{equation}
where $\mathbf{a}$ is the vector potential of $\mathbf{v}$ specified by $\bnabla \cdot
\mathbf{a} = 0$. With this choice of the mean-field coefficients the profile of the toroidal
field clearly deviates from that in the cases considered before. It is less diffused at
midlatitudes and mid radii where convection takes place. This difference can be attributed to the
absence of the $\gamma$-effect. Already in this simple example the isotropic approximation fails
to reproduce the axisymmetric field in satisfactory agreement with corresponding numerical
simulations. This indicates that in general more mean-field coefficients must be taken into account
in order to grasp all relevant dynamo effects. In addition there are deviations of about 50\% in
the amplitudes of the poloidal field.

\subsection{Geodynamo}
\label{sec:6:3}

Consider now again the geodynamo model of section~\ref{sec:5:3}. In figure \ref{fig:geodynamo} the
azimuthally averaged field components resulting from the numerical simulation are shown in
comparison with results given by mean-field modelling. Figure \ref{fig:geodynamo} is organised in
the same way as figure \ref{fig:magnetoconvection} before. That is, azimuthally averaged field
components resulting from a direct numerical simulation have been plotted in the first row, the
second row shows results obtained by corresponding mean-field calculations, the third row
contributes results obtained by mean-field modelling with the coefficients determined in
SOCA, while for the results presented in the last row, the isotropic approximation (\ref{eq8:2})
has been applied. Note that only the direct numerical simulation results in a steady dynamo. All
mean-field models shown in comparison are subcritical and the magnetic fields decay according to
$\bm=\bm_0\exp(-\lambda t)$, where $\bm_0$ denotes the field configuration reached after an
initial transition phase, and the decay rate $\lambda$ is positive in these examples. Therefore,
decay rates rather than amplitudes are compared.

As in the previous example, both mean-field models relying on all 27 mean-field coefficients
(second and third row in figure \ref{fig:geodynamo}) correspond best to the direct numerical
simulation and succeed in reproducing all essential features of the field given in the first row.
However, both mean-field models are slightly subcritical with $\lambda=3.5\,\eta/D^2$ and
$\lambda=6.5\,\eta/D^2$, respectively, which comes along with topological differences in
$\overline{B}_\varphi$. The flux bundles at low latitudes near the outer boundary are more strongly
diffused in the mean-field models. The high diffusion in this region is due to the strong
$\gamma$-effect, which leads to an advection of oppositely oriented mean toroidal fields towards
the equator, resulting in large gradients.

Although SOCA is, strictly speaking, not justified anymore, the resulting mean field components in
the third row are remarkably similar to those obtained by mean-field modelling without applying
SOCA (second row). For an explanation we refer again to the scaling argument given above in the
context of the magnetoconvection example.

Again, mean-field coefficients in the isotropic approximation do not lead to reliable results
anymore, as can be seen from the last row in figure \ref{fig:geodynamo}. There are not only
differences in the field distribution, but also the decay rate, $\lambda\approx 75\,\eta/D^2$, is
drastically high.

\begin{figure}[t]
    \centering\includegraphics[scale=0.8,angle=90]{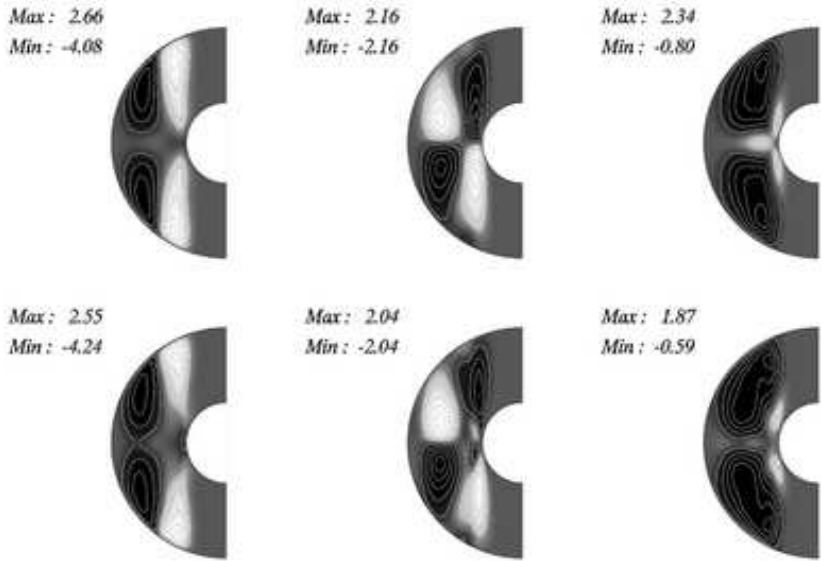}
    \caption{Comparison between electromotive forces in the
      magnetoconvection example. First row:
      $\mathcal{E}^\mathrm{DNS}_r,\mathcal{E}^\mathrm{DNS}_\vartheta,
      \mathcal{E}^\mathrm{DNS}_\varphi$, second row:
      $\mathcal{E}^\mathrm{MF1}_r,\mathcal{E}^\mathrm{MF1}_\vartheta,
      \mathcal{E}^\mathrm{MF1}_\varphi$. Maxima and minima
      are given in units of $(\eta/D)\,(\varrho\mu\eta\Omega)^{1/2}$.}
    \label{fig:emf}
\end{figure}

\subsection{Limits of the representation of the mean electromotive force}
\label{sec:6:4}

The difficulties of mean-field models in accurately reproducing mean magnetic fields compared to
direct numerical simulations, which arise in the example of the geodynamo model, are due to an
inadequate parametrisation of $\emf$. Let us compare $\emf^{\mathrm{DNS}}$, that is $\emf$ as
immediately extracted from the direct numerical simulation, with $\emf^{\mathrm{MF1}}$ defined by
$\mathcal{E}^\mathrm{MF1}_\kappa=\tilde{a}_{\kappa\lambda}\overline{B}_\lambda
+\tilde{b}_{\kappa\lambda r}\,\partial\overline{B}_\lambda/\partial r +
1/r\,\tilde{b}_{\kappa\lambda\vartheta}\,\partial\overline{B}_\lambda/\partial\vartheta$. For the
computation of $\emf^\mathrm{MF1}$, both $\bm$ and its gradient have also been taken from the
direct numerical simulation. We further define the quantity
\begin{equation}
\delta\mathcal{E}=\frac{<|\emf^\mathrm{DNS}-\emf^\mathrm{MF1}|>}{<|\emf^\mathrm{DNS}|>}
\label{eq8:6}
\end{equation}
where $<\cdots>$ means spatial averaging.

Consider first the magnetoconvection example. Figure \ref{fig:emf} shows that $\emf^{\mathrm{DNS}}$
and $\emf^{\mathrm{MF1}}$ are in reasonable agreement. We find $\delta\mathcal{E}\approx 0.28$.
Therefore, a parametrisation of $\emf$ considering no higher than first-order derivatives of
$\bm$ is adequate in this example.

In contrast, this is no longer true for the geodynamo model. In this case $\delta\mathcal{E}>4$
has been found, indicating that a parametrisation according to (\ref{eq3:12}) no longer describes
the actual $\emf$ reasonably well. The  assumption of a sufficiently weak spatial variation of
$\bm$, which is needed to truncate the series expansion of $\emf$ in (\ref{eq3:12}), breaks down.
In fact, higher order derivatives of $\bm$ become large and spoil the representation
(\ref{eq3:12}) of $\emf$ in a rather uncontrolled manner.

This finding is consistent with the following observations. The power spectrum of the radially
averaged mean magnetic field in the example of the geodynamo model exhibits a peak at $l=4$
containing 15\% of the total power. In contrast, the corresponding spectrum in  the
magnetoconvection example does not possess noticeable contributions for $l>2$, suggesting that
the mean-field is indeed smoother in this example. Furthermore, the magnetoconvection simulation
has been repeated with more complicated imposed toroidal fields in order to trigger steeper
gradients in the mean field. In this way, it is indeed possible to destroy the close match between
$\emf^\mathrm{DNS}$ and $\emf^\mathrm{MF1}$ as seen in figure~\ref{fig:emf}.

\subsection{Significance of mean-field coefficients}
\label{sec:6:5}

In order to investigate the significance of the various mean-field coefficients we carried out a
number of test calculations with different sets of mean-field coefficients for the geodynamo case.
As already presented in section~6.3, the inclusion of all $\tilde{a}_{\kappa\lambda}$ and
$\tilde{b}_{\kappa\lambda\nu}$ leads to the decay rate of $\lambda=3.5\,\eta/D^2$.

In one series of calculations we only used the $\tilde{a}_{\kappa\lambda}$ and disregarded the
$\tilde{b}_{\kappa\lambda\nu}$, increased however the molecular diffusivity to $\Pm=1$. With all
$\tilde{a}_{\kappa\lambda}$ the decay rate of the dominant dipolar mode is $\lambda=2.1\,\eta/D^2$.
Using only the diagonal $\tilde{a}_{\kappa\kappa}$ results in $\lambda=-4.8\,\eta/D^2$. Besides the
diagonal terms, $\tilde{a}_{r\varphi}$ and $\tilde{a}_{\vartheta\varphi}$ are most important. They
provide a strong $\gamma$-effect, which as already stated above, acts to expel flux from the
central dynamo region and results in a decay with $\lambda=4.9\,\eta/D^2$. As seen in section
\ref{sec:6:3} by comparing with the numerical simulations, the $\gamma$-effect is crucial for the
geodynamo. Furthermore it leads to a preference of dipolar modes compared to quadrupolar modes,
that is, modes being antisymmetric or symmetric, respectively, about the equatorial plane.

With all $\tilde{a}_{\kappa\lambda}$ coefficients included, we next studied the influence of the
$\tilde{b}_{\kappa\lambda\nu}$ coefficients, now again for a molecular diffusivity described by
$\Pm=5$. With the diagonal terms only, diffusion is significantly enhanced and the resulting dynamo
solution is markedly subcritical with $\lambda=15\,\eta/D^2$. However, not all components of
$\tilde{b}_{\kappa\lambda\nu}$ are conducive to the turbulent diffusion. If in addition all
coefficients which contribute to the $\delta$-effect are considered the decay rate decreases to
$\lambda=3.0\,\eta/D^2$ and the solution gets close to the case where all coefficients are
included. This might be due to a constructive action of the combination of $\delta$-effect and some
kind of mean shear occurring with $\vm$ or $\boldsymbol{\gamma}$, which, if stronger, could lead to
a dynamo even in the absence of the $\alpha$-effect. As explained in appendix~\ref{app:2}, the
$\delta$-effect has no direct influence on the energy stored in the mean magnetic field.

Altogether, the test calculations suggest a large set of mean-field coefficients to be considered
in order to achieve reasonable agreement between mean-field models and numerical simulations. These
are the $\boldsymbol{\alpha}$ and $\boldsymbol{\gamma}$ terms as well as the $\boldsymbol{\beta}$
and $\boldsymbol{\delta}$ terms. The $\boldsymbol{\kappa}$ terms, on the other hand, seem not to be
of importance.

\section{Conclusions}
\label{sec:7}

The knowledge of the mean-field coefficients is decisive in order to analyse and to model dynamo
action in many astrophysical bodies. In this paper, two approaches to determine mean-field
coefficients have been developed. While the numerical approach does not use intrinsic
approximations, the analytical approach is based on the second-order correlation approximation.

The mean-field view is applied to two examples: a simulation of rotating magnetoconvection and of a
quasi-stationary geodynamo. In both examples similar processes take place: the mutual generation of
poloidal and toroidal magnetic fields by an $\alpha$-effect, flux expulsion from the dynamo
region due to a $\gamma$-effect, and a strong turbulent diffusion, which might be moderated by a
$\delta$-effect.

The calculation of mean-field coefficients provides insight into the reliability of frequently
applied approximations in the framework of mean-field theory. Most dubious among them is the
reduction of the $\alpha$-tensor to an isotropic tensor, which leaves dominating non-diagonal
components unconsidered. A further, important simplification is the second-order correlation
approximation. It typically leads to overestimated amplitudes of mean-field coefficients, whereas
their profiles are  rather unaffected.

The mean-field picture of geodynamo models is completed by the simulation of axisymmetric fields by
means of a mean-field model, involving all mean-field coefficients determined. Test calculations
with different sets of mean-field coefficients confirm their relation to the above mentioned dynamo
processes. In  addition, a comparison with azimuthally averaged fields resulting from direct
numerical simulations reveals their significance. In the magnetoconvection and the geodynamo
example considered here, the match between direct numerical simulations and mean-field simulations
is good only if a large number of mean-field coefficients are involved which contribute to
$\boldsymbol{\alpha}$, $\boldsymbol{\gamma}$, $\boldsymbol{\beta}$ and
$\boldsymbol{\delta}$. The application of corresponding mean-field coefficients derived in the
second-order correlation approximation leads to similar results.

The reliability of mean-field models relies on a proper parametrisation of $\emf$ in terms of the
mean magnetic field. In the magnetoconvection example the traditional representation of the mean
electromotive force considering no higher than first order derivatives of the mean magnetic field
is valid. However, already in the geodynamo example it seems no longer to be justified and the
assumption of scale separation is not fulfilled. This limits the applicability of some
approximations commonly used within the mean-field theory. Nonetheless, even in the geodynamo
example the spatial structure of the axisymmetric fields obtained by mean-field modelling
corresponds roughly with that of the azimuthal averages extracted from the corresponding direct
numerical simulation.

\appendix
\section{Approach II -- derivations and further results}
\label{app:1}

We first determine $\mathbf{b}$ so that it satisfies equation (\ref{eq4:10}) inside the fluid shell
and continues as a potential field in its inner and outer surroundings. We represent
$\mathbf{b}$ in the same form as $\mathbf{v}$ in (\ref{eq4:12}) by writing
\begin{equation}
\mathbf{b} = - \bnabla \times (\mathbf{r} \times \bnabla \sigma) - \mathbf{r} \times
\bnabla \tau \label{eqa:2}
\end{equation}
and expanding the scalars $\sigma$ and $\tau$ according to
\begin{equation}
\sigma = \sum_{l,m} \, \sigma_l^m (r) Y_l^m (\vartheta, \varphi) \, , \quad
    \tau = \sum_{l,m} \,  \tau_l^m (r) Y_l^m (\vartheta, \varphi) \, ,
\label{eqa:4}
\end{equation}
where $\sigma_l^m (r)$ and $\tau_l^m (r)$ are complex functions satisfying $\sigma_l^{m *} =
\sigma_l^{-m}$ and $\tau_l^{m *} = \tau_l^{-m}$, but $\sigma_l^0 = \tau_l^0 = 0$, and $Y_l^m
(\vartheta, \varphi)$ stands for spherical harmonics as explained above.

We note that (\ref{eqa:2}) implies
\begin{equation}
L^2 \sigma = - \mathbf{r} \cdot \mathbf{b} \, , \quad
    L^2 \tau = - \mathbf{r} \cdot (\bnabla \times \mathbf{b}) \, ,
\label{eqa:6}
\end{equation}
where
\begin{equation}
L^2 f = \frac{1}{\sin \vartheta} \frac{\partial}{\partial \vartheta}
    \left( \sin \vartheta \, \frac{\partial f}{\partial \vartheta} \right)
    + \frac{1}{\sin^2 \vartheta} \frac{\partial^2 f}{\partial \varphi^2} \, .
\label{eqa:7}
\end{equation}
Of course, if $\mathbf{b}$ in (\ref{eqa:2}) is replaced, e.g., by $\nabla^2 \mathbf{b}$,
(\ref{eqa:6}) applies with the same replacement. We further note that the $Y_l^m$ satisfy the
eigenvalue equation
\begin{equation}
L^2 Y_l^m + l (l + 1) Y_l^m = 0
\label{eqa:9}
\end{equation}
and the orthogonality relation
\begin{equation}
\int  Y_l^m (\vartheta, \varphi) \, Y_{l'}^{m' *} (\vartheta, \varphi) \, \mbox{d} \Omega
    = \frac{4 \pi (l + |m|)!}{(2 l + 1)(l - |m|)!} \delta_{l l'} \delta_{m m'} \, ,
\label{eqa:8}
\end{equation}
where $\mbox{d} \Omega$ means $\sin \vartheta \, \mbox{d} \vartheta \, \mbox{d} \varphi$ and the
integration is over all $\vartheta$ and $\varphi$ of the full solid angle. Here Ferrer's
normalization of the associated Legendre polynomials $P_l^m$ has been adopted.

Using these relations equation (\ref{eq4:10}) can be reduced to
\begin{equation}
\frac{1}{r} \frac{\mbox{d}^2}{\mbox{d} r^2} \big( r \sigma^m_l \big)
   - \frac{l(l+1)}{r^2} \sigma^m_l = F^m_l \, , \quad
\frac{1}{r} \frac{\mbox{d}^2}{\mbox{d} r^2} \big( r \tau^m_l \big)
   - \frac{l(l+1)}{r^2} \tau^m_l = G^m_l \, ,
\label{eqa:10}
\end{equation}
applying in the shell $r_i < r < r_0$, and
\begin{eqnarray}
&& F^m_l = \frac{(2l+1)(l-|m|)!}{4 \pi \eta l(l+1)(l+|m|)!}
   \int (\mathbf{v} \times \bm) \cdot (\mathbf{r} \times\bnabla
   Y^{m*}_l) \, \mbox{d}\Omega
\nonumber\\
&& G^m_l = \frac{(2l+1)(l-|m|)!}{4 \pi \eta l(l+1)(l+|m|)!}
   \int (\bnabla \times (\mathbf{v} \times \bm))
   \cdot (\mathbf{r} \times \bnabla Y^{m*}_l) \, \mbox{d}\Omega \, ,
\label{eqa:12}
\end{eqnarray}
again with integrations over the full solid angle. The continuation of $\mathbf{b}$ as a
potential field in the regions inside and outside the conducting shell requires
\begin{equation}
\frac{\mbox{d} \sigma^m_l}{\mbox{d} r} - \frac{l}{r} \sigma^m_l = \tau^m_l  =  0
   \quad \mbox{at} \quad r = r_i \, , \quad
   \frac{\mbox{d} \sigma^m_l}{\mbox{d} r} + \frac{l+1}{r} \sigma^m_l = \tau^m_l  = 0
   \quad \mbox{at} \quad r = r_0 \, .
\label{eqa:14}
\end{equation}

The solutions of (\ref{eqa:10}) satisfying (\ref{eqa:14}) can be written in the form
\begin{equation}
\sigma^m_l = - \int_{r_i}^{r_0} f_l (r, r') F^m_l (r') {r'}^2 \mbox{d} r' \, , \quad
    \tau^m_l = - \int_{r_i}^{r_0} g_l (r, r') G^m_l (r') {r'}^2 \mbox{d} r'
\label{eqa:16}
\end{equation}
with Green's functions  $f_l$ and $g_l$ defined by
\begin{equation}
  f_l (r, r') = \frac{1}{(2l+1)r} \left\{
    \begin{array}{cl}
      \left(\displaystyle{\frac{r'}{r}}\right)^l\,, &
      r' \leq r\\
      &\\
      \left(\displaystyle{\frac{r}{r'}}\right)^{l+1}\,, &
      r' \geq r
    \end{array}
  \right.
  \label{eqa:18}
\end{equation}
and
\begin{equation}
  g_l (r, r') = \left\{
    \begin{array}{rl}
      \left(1-\left(\displaystyle{\frac{r_i}{r_0}}
      \right)^{2l+1}\right)^{-1}\,
      \left[
      f_l(r,r')-\left(\displaystyle{\frac{r}{r_0}}\right)^l
      f_l(r_0,r')\right.\\&\\\left.-
      \left(\displaystyle{\frac{r_i}{r}}\right)^{l+1}f_l(r_i,r')
      +\displaystyle{\frac{r^l r_i^{l+1}}
      {r_0^{2l+1}}f_l(r_i,r')}
      \right],& r' \leq r\\
      &\\
      \left(1-\left(\displaystyle{\frac{r}{r_0}}
      \right)^{2l+1}\right)^{-1}\,
      \left[
      f_l(r,r')-\left(\displaystyle{\frac{r}{r_0}}\right)^l
      f_l(r_0,r')\right.\\&\\\left.-
      \left(\displaystyle{\frac{r_i}{r}}\right)^{l+1}f_l(r_i,r')
      +\displaystyle{\frac{r_i^{2l+1}}{r^{l+1}r_0^l}f_l(r_0,r')}
      \right],& r' \geq r \,.
    \end{array}
  \right.
  \label{eqa:20}
\end{equation}
Indeed it becomes clear by inserting of (\ref{eqa:16}) that (\ref{eqa:10}) is satisfied.
Since (\ref{eqa:18}) and (\ref{eqa:20}) imply
\begin{eqnarray}
\frac{\partial f_l (r, r')}{\partial r} - \frac{l}{r} f_l (r, r')
  &=&  g_l (r, r') = 0 \quad \mbox{at} \quad r = r_i
\nonumber\\
\frac{\partial f_l (r, r')}{\partial r} + \frac{l+1}{r} f_l (r, r')
  &=&  g_l (r, r') = 0 \quad \mbox{at} \quad r = r_0 \, ,
\label{eqa:22}
\end{eqnarray}
(\ref{eqa:16}) also satisfies (\ref{eqa:14}).

Inserting now the representation of $\mathbf{v}$ as given by (\ref{eq4:12}) and
(\ref{eq4:14}) into the relation for $F_l^m$ given by (\ref{eqa:12}) and then the result for
$F^m_l$ into the expression for $\sigma^m_l$ in (\ref{eqa:16}), we arrive at
\begin{eqnarray}
\sigma^m_l(r)&=& \frac{(2l+1)(l-|m|)!}{4 \pi \eta l(l+1)(l+|m|)!}\nonumber\\
&&\sum_{l'} \int f_l (r, r')
    \bigg\{ \Big[ {\hat{\phi}}^m_{l'} (r') R^m_{l' l} (\vartheta')
    \nonumber\\
&&  - \mbox{i} m \psi^m_{l'} (r')
    \Big(Q^m_{l' l} (\vartheta') + Q^m_{l l'} (\vartheta')\Big)\Big/\sin \vartheta' \Big]
    {\overline{B}}_{r} (r', \vartheta')
\nonumber\\
&& - \frac{l'(l'+1)}{r'} \phi^m_{l'} (r') Q^m_{l' l} (\vartheta')
    {\overline{B}}_{\vartheta} (r', \vartheta')
\nonumber\\
&& + \mbox{i} m \frac{l'(l'+1)}{r'} \phi^m_{l'} (r')
    \Big(P^m_{l' l} (\vartheta') / \sin \vartheta'\Big)
    {\overline{B}}_{\varphi} (r', \vartheta') \bigg\} \mbox{d} v' \, .
\label{eqa:30}
\end{eqnarray}
The integration is over the whole fluid shell. When proceeding analogously with the relation for
$G_l^m$ in (\ref{eqa:12}), the result for $G^m_l$ contains derivatives of $\overline{B}_r$,
$\overline{B}_\vartheta$, and $\overline{B}_\varphi$ with respect to $r$ and
$\vartheta$. We may remove them  by means of integration by parts. In this way we find
\begin{eqnarray}
\tau^m_l (r) &=& \frac{(2l+1)(l-|m|)!}{4 \pi \eta l(l+1)(l+|m|)!}
\sum_{l'} \int
\bigg\{ \tilde{g}_l (r, r')\nonumber \\
&& \times\Big[\mbox{i} m {\hat{\phi}}^m_{l'} (r')
    \Big(Q^m_{l' l} (\vartheta') + Q^m_{l l'} (\vartheta')\Big)\Big/\sin \vartheta'
   - \psi^m_{l'} (r') R^m_{l' l} (\vartheta') \Big]
    {\overline{B}}_{r} (r', \vartheta')
\nonumber\\
&& - \frac{1}{r'} \Big[ l(l+1) g_l (r, r')
    \Big(\mbox{i} m {\hat{\phi}}^m_{l'} (r') P^m_{l' l}
    (\vartheta') / \sin \vartheta'
    - \psi^m_{l'} (r') Q^m_{l l'} (\vartheta')\Big)\nonumber\\
&& + \mbox{i} m l'(l'+1) \tilde{g}_l (r, r') \, \phi^m_{l'} (r')
    P^m_{l' l} (\vartheta') / \sin \vartheta' \Big]
    {\overline{B}}_{\vartheta} (r', \vartheta')\nonumber\\
&& + \frac{1}{r'} \Big[ l(l+1) g_l (r, r')
    \Big({\hat{\phi}}^m_{l'} (r') Q^m_{l l'} (\vartheta')
    + \mbox{i} m \psi^m_{l'} (r') P^m_{l l'} (\vartheta') / \sin \vartheta'\Big)
\nonumber\\
&& - l'(l'+1) \tilde{g}_l (r, r') \, \phi^m_{l'} (r')
     Q^m_{l' l} (\vartheta') \Big]
     {\overline{B}}_{\varphi} (r', \vartheta') \bigg\} \mbox{d}v'
\label{eqa:32}
\end{eqnarray}
with $\tilde{g}_l (r, r') = (1 / r') \, \partial \big( r' g_l (r,r')\big)/ \partial r'$.

The relations (\ref{eqa:2}) and (\ref{eqa:4}) together with (\ref{eqa:30}) and (\ref{eqa:32})
represent the wanted solution of the equation (\ref{eq4:10}) for $\mathbf{b}$.

Let us now proceed to $\emf = \vxb$. Expressing $\mathbf{v}$ according to (\ref{eq4:12}) and
(\ref{eq4:14}) and $\mathbf{b}$ according to (\ref{eqa:2}) and (\ref{eqa:4}) we find
\begin{eqnarray}
{\mathcal{E}}_r &=& - 2 \sum_{l, l' \, ; \,  m > 0}
   \bigg[ \Real \Big({\hat{\phi}}_{l'}^{m*} \tau_l^m - \psi_{l'}^{m*} {\hat{\sigma}}_l^m\Big) \, R^m_{l' l}
\nonumber\\
    && + m \,\mbox{Im} \Big({\hat{\phi}}_{l'}^{m*} {\hat{\sigma}}_l^m
    + \psi_{l'}^{m*} \tau_l^m\Big) \, \big(Q^m_{l' l} +
    Q^m_{l l'}\big) / \sin \vartheta \bigg]
\nonumber\\
{\mathcal{E}}_\vartheta &=& + \frac{2}{r} \sum_{l, l' \, ; \,  m > 0}
   \bigg[ l'(l'+1)\Real \Big(\phi_{l'}^{m*} \tau_l^m\Big) \, Q^m_{l' l}
   - l(l+1) \Real \Big(\psi_{l'}^{m*} \sigma_l^m\Big) \, Q^m_{l l'}\nonumber\\
   && + m \, \bigg( l(l+1)\mbox{Im} \Big({\hat{\phi}}_{l'}^{m*} \sigma_l^m\Big)
   + l'(l'+1) \mbox{Im} \Big(\phi_{l'}^{m*} {\hat{\sigma}}_l^m\Big) \bigg) \,
   P^m_{l' l} / \sin \vartheta \bigg]
\label{eqa:40}\\
{\mathcal{E}}_\varphi &=& - \frac{2}{r} \sum_{l, l' \, ; \,  m > 0}
   \bigg[ l(l+1)\Real \Big({\hat{\phi}}_{l'}^{m*} \sigma_l^m\Big) \, Q^m_{l l'}
   - l'(l'+1) \Real \Big(\phi_{l'}^{m*} {\hat{\sigma}}_l^m\Big) \, Q^m_{l' l}
\nonumber\\
   && + m \, \bigg( l'(l'+1) \mbox{Im} \Big(\phi_{l'}^{m*} \tau_l^m\Big)
   + l(l+1) \mbox{Im} \Big(\psi_{l'}^{m*} \sigma_l^m\Big) \bigg) \,
   P^m_{l' l} / \sin \vartheta \bigg] \, .
\nonumber
\end{eqnarray}
The $\phi_l^m$, $\psi_l^m$, $\sigma_l^m$ and $\tau_l^m$ depend of course on $r$. The $P_{l'l}^m$,
$Q_{l'l}^m$ and $R_{l'l}^m$ are functions of $\vartheta$ defined by
\begin{eqnarray}
P^{m}_{l' l} & = & P^{|m|}_{l'} (\cos \vartheta) P^{|m|}_l (\cos
\vartheta)
\nonumber\\
Q^{m}_{l' l} & = &  P^{|m|}_{l'} (\cos \vartheta)
   \frac{\mbox{d} P^{|m|}_l (\cos \vartheta)}{\mbox{d} \vartheta}
\label{eqa:42}\\
R^m_{l' l} & = & \frac{\mbox{d} P^{|m|}_{l'} (\cos \vartheta)}{\mbox{d} \vartheta}
   \frac{\mbox{d} P^{|m|}_l (\cos \vartheta)}{\mbox{d} \vartheta}
   + \frac{m^2}{\sin^2 \vartheta} P^{|m|}_{l'} (\cos \vartheta)
   P^{|m|}_{l} (\cos \vartheta) \, .
\nonumber
\end{eqnarray}

We may now insert the results (\ref{eqa:30}) and  (\ref{eqa:32}) for $\sigma_l^m$ and $\tau_l^m$ into
(\ref{eqa:40}). Then the ${\mathcal{E}}_\kappa$ take indeed the form (\ref{eq4:50}). Unfortunately
the $K_{\kappa \lambda}$ are rather complex expressions. As an example we mention
\begin{eqnarray}
K_{rr} (r, \vartheta; r', \vartheta') &=& \frac{1}{2 \pi \eta}
    \sum_{l, l', l'' \, ; \ m > 0}
    \frac{(2l+1)(l-|m|)!}{l(l+1)(l+|m|)!}
\nonumber\\
&&\times\Bigg\{ \bigg[ {\hat{f}}_l (r, r') \,
\mbox{Im}\Big(\psi^{m*}_{l'}(r) \psi^m_{l''}(r')\Big)
     + {\tilde{g}}_l (r, r') \, \mbox{Im}\Big({\hat{\phi}}^{m*}_{l'}(r)
     {\hat{\phi}}^m_{l''}(r')\Big) \bigg]
\nonumber\\
&& \hspace*{5mm} \times m \, R^m_{l' l}(\vartheta) \,
     \Big(Q^m_{l'' l}(\vartheta') + Q^m_{l l''}(\vartheta')\Big)\Big/\sin \vartheta'  \,
\nonumber\\
&& + \bigg[ {\hat{f}}_l (r, r') \, \Real \Big({\hat{\phi}}^{m*}_{l'}(r) \psi^m_{l''}(r')\Big)
     - {\tilde{g}}_l (r, r') \, \Real \Big(\psi^{m*}_{l'}(r) {\hat{\phi}}^m_{l''}(r')\Big) \bigg]
\nonumber\\
&& \hspace*{5mm} \times m^2 \, \Big(Q^m_{l' l}(\vartheta) + Q^m_{l l'}(\vartheta)\Big) \,
     \Big(Q^m_{l'' l}(\vartheta') + Q^m_{l l''}(\vartheta')\Big)\Big/(\sin\vartheta\sin\vartheta')
\nonumber\\
&& + \bigg[ {\hat{f}}_l (r, r') \, \Real \Big(\psi^{m*}_{l'}(r)
     {\hat{\phi}}^m_{l''}(r')\Big)
     + {\tilde{g}}_l (r, r') \, \Real \Big({\hat{\phi}}^{m*}_{l'}(r) \psi^m_{l''}(r')\Big) \bigg]
\nonumber\\
&& \hspace*{5mm} \times R^m_{l' l}(\vartheta) \, R^m_{l'' l}(\vartheta')
\nonumber\\
&& - \bigg[ {\hat{f}}_l (r, r') \, \mbox{Im}\Big({\hat{\phi}}^{m*}_{l'}(r)
     {\hat{\phi}}^m_{l''}(r')\Big)
     - {\tilde{g}}_l (r, r') \, \mbox{Im}\Big(\psi^{m*}_{l'}(r) \psi^m_{l''}(r')\Big) \bigg]
\nonumber\\
&& \hspace*{5mm} \times m\,\Big(Q^m_{l' l}(\vartheta) + Q^m_{l l'}(\vartheta)\Big)\,R^m_{l''l}(\vartheta')/\sin\vartheta  \Bigg\}
\label{eqa:44}
\end{eqnarray}
where $\hat{f}_l (r, r') = (1/r) \, \partial\big(r f_l(r,r')\big)/\partial r$.

Using now ({\ref{eq4:51}) and the orthogonality relations
\begin{eqnarray}
&& \int_0^\pi P^m_{l' l} (\vartheta) \sin \vartheta \mbox{d} \vartheta
    = \frac{2(l+|m|)!}{(2l+1)(l-|m|)!} \delta_{l' l}
\nonumber\\
&& \int_0^\pi \big( Q^m_{l' l}(\vartheta) + Q^m_{l l'}(\vartheta) \big) \mbox{d} \vartheta
    = 0
\label{eqa:48}\\
&& \int_0^\pi R^m_{l' l} (\vartheta) \sin \vartheta \mbox{d} \vartheta
    = \frac{2l(l+1)(l+|m|)!}{(2l+1)(l-|m|)!} \delta_{l' l}
\nonumber
\end{eqnarray}
we find $\tilde{a}_{rr}$ as given by (\ref{eq4:58}).

In the same way all other $\tilde{a}_{\kappa \lambda}$ can be calculated. Unfortunately in the
cases of $\tilde{a}_{\kappa \vartheta}$ and $\tilde{a}_{\kappa \varphi}$ the integration over
$\vartheta$ can not be carried out with taking benefit of orthogonality relations like
(\ref{eqa:48}), and the need of numerical integrations remains. The results read
\begin{eqnarray}
\tilde{a}_{\vartheta r}(r,\vartheta) &=& \frac{2}{\eta r}\sum_{l,l'\,;\,m>0} \Bigg\{
-l'(l'+1)\int_{r_i}^{r_0} \tilde{g}_l(r,r')
   \mbox{Re}\Big(\phi^{m*}_{l'}(r)\psi^m_{l}(r')\Big)
   \,{r'}^2\mbox{d}r' \; Q^m_{l'l}(\vartheta) \nonumber\\
&& -\;l(l+1)\int_{r_i}^{r_0} f_l(r,r')
   \mbox{Re}\Big(\psi^{m*}_{l'}(r)\hat{\phi}^m_l(r')\Big)
   \,{r'}^2\mbox{d}r' \; Q^m_{ll'}(\vartheta) \nonumber\\
&& +\;m\int_{r_i}^{r_0}\bigg[l'(l'+1)\hat{f}_l(r,r')
   \mbox{Im}\Big(\phi^{m*}_{l'}(r)\hat{\phi}^m_l(r')\Big)\nonumber\\
&& +l(l+1)f_l(r,r')
   \mbox{Im}\Big(\hat{\phi}^{m*}_{l'}(r)\hat{\phi}^m_l(r')\Big)\bigg]
   \,{r'}^2\mbox{d}r' \; P^m_{l'l}(\vartheta)/\sin\vartheta \Bigg\}
\label{eqa:52}
\end{eqnarray}

\begin{eqnarray}
\tilde{a}_{\varphi r}(r,\vartheta) &=& \frac{2}{\eta r}\sum_{l,l'\,;\,m>0} \Bigg\{
\,l'(l'+1)\int_{r_i}^{r_0} \hat{f}_l(r,r')
   \mbox{Re}\Big(\phi^{m*}_{l'}(r)\hat{\phi}^m_{l}(r')\Big)
   \,{r'}^2\mbox{d}r' \; Q^m_{l'l}(\vartheta) \nonumber\\
&& -\;l(l+1)\int_{r_i}^{r_0} f_l(r,r')
   \mbox{Re}\Big(\hat{\phi}^{m*}_{l'}(r)\hat{\phi}^m_l(r')\Big)
   \,{r'}^2\mbox{d}r' \; Q^m_{ll'}(\vartheta) \nonumber\\
&& +\;m\int_{r_i}^{r_0}\bigg[l'(l'+1)\tilde{g}_l(r,r')
   \mbox{Im}\Big(\phi^{m*}_{l'}(r)\psi^m_l(r')\Big) \nonumber\\
&& -l(l+1)f_l(r,r')
   \mbox{Im}\Big(\psi^{m*}_{l'}(r)\hat{\phi}^m_l(r')\Big)\bigg]
   \,{r'}^2\mbox{d}r' \; P^m_{l'l}(\vartheta)/\sin\vartheta \Bigg\}
\label{eqa:54}
\end{eqnarray}

\begin{eqnarray}
\tilde{a}_{r\vartheta}(r,\vartheta) &=& \frac{1}{\eta}\sum_{l,l',l''\,;\,m>0}
\frac{(2l+1)(l-|m|)!}{l(l+1)(l+|m|)!} \nonumber\\
&& \times\Bigg\{ R^m_{l'l}(\vartheta)
   \Bigg[ -m\int_{r_i}^{r_0} \bigg\{ l(l+1)g_l(r,r')
   \mbox{Im}\Big(\hat{\phi}^{m*}_{l'}(r)\hat{\phi}^m_{l''}(r')\Big) \nonumber\\
&& +l''(l''+1)\tilde{g}_l(r,r')
   \mbox{Im}\Big(\hat{\phi}^{m*}_{l'}(r)\phi^m_{l''}(r')\Big) \bigg\}
   \,r'\mbox{d}r'
   \int_0^\pi P^m_{l''l}(\vartheta')\mbox{d}\vartheta' \nonumber\\
&& -l(l+1)\int_{r_i}^{r_0} g_l(r,r')
   \mbox{Re}\Big(\hat{\phi}^{m*}_{l'}(r)\psi^m_{l''}(r')\Big)\,r'\mbox{d}r'
   \int_0^\pi Q^m_{ll''}(\vartheta')\sin\vartheta'\mbox{d}\vartheta' \nonumber\\
&& -l''(l''+1)\int_{r_i}^{r_0} \hat{f}_l(r,r')
   \mbox{Re}\Big(\psi^{m*}_{l'}(r)\phi^m_{l''}(r')\Big)\,r'\mbox{d}r'
   \int_0^\pi Q^m_{l''l}(\vartheta')\sin\vartheta'\mbox{d}\vartheta' \Bigg] \nonumber\\
&& +m\Big(Q^m_{l'l}(\vartheta)+Q^m_{ll'}(\vartheta)\Big)\big/\sin\vartheta
   \Bigg[ m\int_{r_i}^{r_0} \bigg\{ l(l+1)g_l(r,r')
   \mbox{Re}\Big(\psi^{m*}_{l'}(r)\hat{\phi}^m_{l''}(r')\Big) \nonumber\\
&& +l''(l''+1)\tilde{g}_l(r,r')
   \mbox{Re}\Big(\psi^{m*}_{l'}(r)\phi^m_{l''}(r')\Big) \bigg\}
   \,r'\mbox{d}r'
   \int_0^\pi P^m_{l''l}(\vartheta')\mbox{d}\vartheta' \nonumber\\
&& -l(l+1)\int_{r_i}^{r_0} g_l(r,r')
   \mbox{Im}\Big(\psi^{m*}_{l'}(r)\psi^m_{l''}(r')\Big)\,r'\mbox{d}r'
   \int_0^\pi Q^m_{ll''}(\vartheta')\sin\vartheta'\mbox{d}\vartheta' \nonumber\\
&& +l''(l''+1)\int_{r_i}^{r_0} \hat{f}_l(r,r')
   \mbox{Im}\Big(\hat{\phi}^{m*}_{l'}(r)\phi^m_{l''}(r')\Big)\,r'\mbox{d}r'
   \nonumber\\
&& \times\int_0^\pi Q^m_{l''l}(\vartheta')\sin\vartheta'\mbox{d}\vartheta'
   \Bigg]
\Bigg\}
\label{eqa:56}
\end{eqnarray}

\begin{eqnarray}
\tilde{a}_{\vartheta\vartheta}(r,\vartheta) &=& \frac{1}{\eta r}\sum_{l,l',l''\,;\,m>0}
\frac{(2l+1)(l-|m|)!}{l(l+1)(l+|m|)!} \nonumber\\
&& \times\Bigg\{ \,l'(l'+1)Q^m_{l'l}(\vartheta)
   \Bigg[ m\int_{r_i}^{r_0} \bigg\{ l(l+1)g_l(r,r')
   \mbox{Im}\Big(\phi^{m*}_{l'}(r)\hat{\phi}^m_{l''}(r')\Big) \nonumber\\
&& +\;l''(l''+1)\tilde{g}_l(r,r')
   \mbox{Im}\Big(\phi^{m*}_{l'}(r)\phi^m_{l''}(r')\Big) \bigg\}
   \,r'\mbox{d}r'
   \int_0^\pi P^m_{l''l}(\vartheta')\mbox{d}\vartheta' \nonumber\\
&& +\;l(l+1)\int_{r_i}^{r_0} g_l(r,r')
   \mbox{Re}\Big(\phi^{m*}_{l'}(r)\psi^m_{l''}(r')\Big)\,r'\mbox{d}r'
   \int_0^\pi Q^m_{ll''}(\vartheta')\sin\vartheta'\mbox{d}\vartheta' \Bigg]
   \nonumber\\
&& -\;l''(l''+1)
   \int_0^\pi Q^m_{l''l}(\vartheta')\sin\vartheta'\mbox{d}\vartheta'
   \Bigg[ m\int_{r_i}^{r_0} \bigg\{ l(l+1)f_l(r,r')
   \mbox{Im}\Big(\hat{\phi}^{m*}_{l'}(r)\phi^m_{l''}(r')\Big) \nonumber\\
&& +\;l'(l'+1)\hat{f}_l(r,r')
   \mbox{Im}\Big(\phi^{m*}_{l'}(r)\phi^m_{l''}(r')\Big) \bigg\}
   \,r'\mbox{d}r' \; P^m_{l'l}(\vartheta)/\sin\vartheta \nonumber\\
&& -\;l(l+1)\int_{r_i}^{r_0} f_l(r,r')
   \mbox{Re}\Big(\psi^{m*}_{l'}(r)\phi^m_{l''}(r')\Big)\,r'\mbox{d}r'
   \,Q^m_{ll'}(\vartheta) \Bigg]
\Bigg\}
\label{eqa:58}
\end{eqnarray}

\begin{eqnarray}
\tilde{a}_{\varphi\vartheta}(r,\vartheta) &=& \frac{1}{\eta r}\sum_{l,l',l''\,;\,m>0}
\frac{(2l+1)(l-|m|)!}{l(l+1)(l+|m|)!} \nonumber\\
&& \times\Bigg\{ \,ml'(l'+1)P^m_{l'l}(\vartheta)/\sin\vartheta
   \Bigg[ m\int_{r_i}^{r_0} \bigg\{ l(l+1)g_l(r,r')
   \mbox{Re}\Big(\phi^{m*}_{l'}(r)\hat{\phi}^m_{l''}(r')\Big) \nonumber\\
&& +\;l''(l''+1)\tilde{g}_l(r,r')
   \mbox{Re}\Big(\phi^{m*}_{l'}(r)\phi^m_{l''}(r')\Big) \bigg\}
   \,r'\mbox{d}r'
   \int_0^\pi P^m_{l''l}(\vartheta')\mbox{d}\vartheta' \nonumber\\
&& -\;l(l+1)\int_{r_i}^{r_0} g_l(r,r')
   \mbox{Im}\Big(\phi^{m*}_{l'}(r)\psi^m_{l''}(r')\Big)\,r'\mbox{d}r'
   \int_0^\pi Q^m_{ll''}(\vartheta')\sin\vartheta'\mbox{d}\vartheta' \Bigg]
   \nonumber\\
&& +\;l''(l''+1)
   \int_0^\pi Q^m_{l''l}(\vartheta')\sin\vartheta'\mbox{d}\vartheta'
   \nonumber\\
&& \Bigg[ ml(l+1) P^m_{l'l}(\vartheta)/\sin\vartheta
   \int_{r_i}^{r_0} f_l(r,r')
   \mbox{Im}\Big(\psi^{m*}_{l'}(r)\phi^m_{l''}(r')\Big) \,r'\mbox{d}r'
   \nonumber\\
&& l'(l'+1) Q^m_{l'l}(\vartheta) \int_{r_i}^{r_0} \hat{f}_l(r,r')
   \mbox{Re}\Big(\phi^{m*}_{l'}(r)\phi^m_{l''}(r')\Big) \,r'\mbox{d}r'
   \nonumber\\
&& +\;l(l+1) Q^m_{ll'}(\vartheta)           \int_{r_i}^{r_0} f_l(r,r')
   \mbox{Re}\Big(\hat{\phi}^{m*}_{l'}(r)\phi^m_{l''}(r')\Big)\,r'\mbox{d}r'
   \Bigg]
\Bigg\}
\label{eqa:60}
\end{eqnarray}

\begin{eqnarray}
\tilde{a}_{r\varphi}(r,\vartheta) &=& -\;\frac{1}{\eta}\sum_{l,l',l''\,;\,m>0}
\frac{(2l+1)(l-|m|)!}{l(l+1)(l+|m|)!} \nonumber\\
&& \times\Bigg\{ R^m_{l'l}(\vartheta)
   \Bigg[ -\;m\int_{r_i}^{r_0} \bigg\{ l(l+1)g_l(r,r')
   \mbox{Im}\Big(\hat{\phi}^{m*}_{l'}(r)\psi^m_{l''}(r')\Big) \nonumber\\
&& -\;l''(l''+1)\hat{f}_l(r,r')
   \mbox{Im}\Big(\psi^{m*}_{l'}(r)\phi^m_{l''}(r')\Big) \bigg\}
   \,r'\mbox{d}r'
   \int_0^\pi P^m_{l''l}(\vartheta')\mbox{d}\vartheta' \nonumber\\
&& +\;l(l+1)\int_{r_i}^{r_0} g_l(r,r')
   \mbox{Re}\Big(\hat{\phi}^{m*}_{l'}(r)\hat{\phi}^m_{l''}(r')\Big)\,r'\mbox{d}r'
   \int_0^\pi Q^m_{ll''}(\vartheta')\sin\vartheta'\mbox{d}\vartheta' \nonumber\\
&& -\;l''(l''+1)\int_{r_i}^{r_0} \tilde{g}_l(r,r')
   \mbox{Re}\Big(\hat{\phi}^{m*}_{l'}(r)\phi^m_{l''}(r')\Big)\,r'\mbox{d}r'
   \int_0^\pi Q^m_{l''l}(\vartheta')\sin\vartheta'\mbox{d}\vartheta' \Bigg] \nonumber\\
&& +\;m\Big(Q^m_{l'l}(\vartheta)+Q^m_{ll'}(\vartheta)\Big)\big/\sin\vartheta
   \Bigg[ m\int_{r_i}^{r_0} \bigg\{ l(l+1)g_l(r,r')
   \mbox{Re}\Big(\psi^{m*}_{l'}(r)\psi^m_{l''}(r')\Big) \nonumber\\
&& +\;l''(l''+1)\hat{f}_l(r,r')
   \mbox{Re}\Big(\hat{\phi}^{m*}_{l'}(r)\phi^m_{l''}(r')\Big) \bigg\}
   \,r'\mbox{d}r'
   \int_0^\pi P^m_{l''l}(\vartheta')\mbox{d}\vartheta' \nonumber\\
&& +\;l(l+1)\int_{r_i}^{r_0} g_l(r,r')
   \mbox{Im}\Big(\psi^{m*}_{l'}(r)\hat{\phi}^m_{l''}(r')\Big)\,r'\mbox{d}r'
   \int_0^\pi Q^m_{ll''}(\vartheta')\sin\vartheta'\mbox{d}\vartheta' \nonumber\\
&& -\;l''(l''+1)\int_{r_i}^{r_0} \tilde{g}_l(r,r')
   \mbox{Im}\Big(\psi^{m*}_{l'}(r)\phi^m_{l''}(r')\Big)\,r'\mbox{d}r'
\nonumber\\
&& \times\int_0^\pi Q^m_{l''l}(\vartheta')\sin\vartheta'\mbox{d}\vartheta'
   \Bigg]
\Bigg\}
\label{eqa:62}
\end{eqnarray}

\begin{eqnarray}
\tilde{a}_{\vartheta\varphi}(r,\vartheta) &=& \frac{1}{\eta r}\sum_{l,l',l''\,;\,m>0}
\frac{(2l+1)(l-|m|)!}{l(l+1)(l+|m|)!} \nonumber\\
&& \times\Bigg\{ \,l'(l'+1)Q^m_{l'l}(\vartheta)
   \Bigg[ -\;ml(l+1) \int_{r_i}^{r_0} g_l(r,r')
   \mbox{Im}\Big(\phi^{m*}_{l'}(r)\psi^m_{l''}(r')\Big) \,r'\mbox{d}r'
   \int_0^\pi P^m_{ll''}(\vartheta')\mbox{d}\vartheta' \nonumber\\
&& -\;l''(l''+1) \int_{r_i}^{r_0} \tilde{g}_l(r,r')
   \mbox{Re}\Big(\phi^{m*}_{l'}(r)\phi^m_{l''}(r')\Big) \,r'\mbox{d}r'
   \int_0^\pi Q^m_{l''l}(\vartheta')\sin\vartheta'\mbox{d}\vartheta' \nonumber\\
&& +\;l(l+1)\int_{r_i}^{r_0} g_l(r,r')
   \mbox{Re}\Big(\phi^{m*}_{l'}(r)\hat{\phi}^m_{l''}(r')\Big)\,r'\mbox{d}r'
   \int_0^\pi Q^m_{ll''}(\vartheta')\sin\vartheta'\mbox{d}\vartheta' \Bigg]
   \nonumber\\
&& +\;ml''(l''+1)
   \int_0^\pi P^m_{l''l}(\vartheta')\mbox{d}\vartheta'
   \Bigg[ m\int_{r_i}^{r_0} \bigg\{ l(l+1)f_l(r,r')
   \mbox{Re}\Big(\hat{\phi}^{m*}_{l'}(r)\phi^m_{l''}(r')\Big) \nonumber\\
&& +\;l'(l'+1)\hat{f}_l(r,r')
   \mbox{Re}\Big(\phi^{m*}_{l'}(r)\phi^m_{l''}(r')\Big) \bigg\}
   \,r'\mbox{d}r' \; P^m_{l'l}(\vartheta)/\sin\vartheta \nonumber\\
&& +\;l(l+1)\int_{r_i}^{r_0} f_l(r,r')
   \mbox{Im}\Big(\psi^{m*}_{l'}(r)\phi^m_{l''}(r')\Big)\,r'\mbox{d}r'
   \,Q^m_{ll'}(\vartheta) \Bigg]
\Bigg\}
\label{eqa:64}
\end{eqnarray}

\begin{eqnarray}
\tilde{a}_{\varphi\varphi}(r,\vartheta) &=& -\;\frac{1}{\eta r}
\sum_{l,l',l''\,;\,m>0}
\frac{(2l+1)(l-|m|)!}{l(l+1)(l+|m|)!} \nonumber\\
&& \times\Bigg\{ \,ml'(l'+1)P^m_{l'l}(\vartheta)/\sin\vartheta\nonumber\\
&& \times\Bigg[ \;ml(l+1) \int_{r_i}^{r_0} g_l(r,r')
   \mbox{Re}\Big(\phi^{m*}_{l'}(r)\psi^m_{l''}(r')\Big) \,r'\mbox{d}r'
   \int_0^\pi P^m_{ll''}(\vartheta')\mbox{d}\vartheta' \nonumber\\
&& -\;l''(l''+1) \int_{r_i}^{r_0} \tilde{g}_l(r,r')
   \mbox{Im}\Big(\phi^{m*}_{l'}(r)\phi^m_{l''}(r')\Big) \,r'\mbox{d}r'
   \int_0^\pi Q^m_{l''l}(\vartheta')\sin\vartheta'\mbox{d}\vartheta'
   \nonumber\\
&& +\;l(l+1)\int_{r_i}^{r_0} g_l(r,r')
   \mbox{Im}\Big(\phi^{m*}_{l'}(r)\hat{\phi}^m_{l''}(r')\Big)\,r'\mbox{d}r'
   \int_0^\pi Q^m_{ll''}(\vartheta')\sin\vartheta'\mbox{d}\vartheta' \Bigg]
   \nonumber\\
&& +\;ml''(l''+1)
   \int_0^\pi P^m_{l''l}(\vartheta')\mbox{d}\vartheta'\nonumber\\
&& \times\Bigg[ ml(l+1) P^m_{l'l}(\vartheta)/\sin\vartheta \int_{r_i}^{r_0} f_l(r,r')
   \mbox{Re}\Big(\psi^{m*}_{l'}(r)\phi^m_{l''}(r')\Big) \,r'\mbox{d}r'
   \nonumber\\
&& +\;l'(l'+1) Q^m_{l'l}(\vartheta) \int_{r_i}^{r_0} \hat{f}_l(r,r')
   \mbox{Im}\Big(\phi^{m*}_{l'}(r)\phi^m_{l''}(r')\Big) \,r'\mbox{d}r'
   \nonumber\\
&& -\;l(l+1) Q^m_{ll'}(\vartheta) \int_{r_i}^{r_0} f_l(r,r')
   \mbox{Im}\Big(\hat{\phi}^{m*}_{l'}(r)\phi^m_{l''}(r')\Big)\,r'\mbox{d}r'
   \Bigg]
\Bigg\} \,.
\label{eqa:66}
\end{eqnarray}

\section{Mean-field energy balance}
\label{app:2}

Above we have applied the mean-field concept to the induction equation (\ref{eq2:8}). For the
following considerations it is more convenient to start from Maxwell's equations in the
quasi-steady approximation. Their mean-field version reads
\begin{equation}
\bnabla \times \Em = - \frac{\partial \bm}{\partial t} \, , \quad
\bnabla \times \bm = \mu \overline{\mathbf{j}} \, , \quad
\bnabla \cdot \bm = 0 \, .
\label{eqc:2}
\end{equation}
Consider a finite fluid body embedded in free space and assume that there are no causes of $\bm$ at
infinity. Then, by standard reasoning, the relation
\begin{equation}
\frac{d}{dt}\int_\infty \frac{\bm^2}{2\mu}dv=-\int_\mathcal{V} \jm\cdot\Em\,dv \label{eqc:4}
\end{equation}
can be derived. The integral on the left-hand side is over all infinite space and thus gives the
total magnetic energy stored in the mean magnetic field whereas that on the right-hand side is over
the fluid body only. Consider now the mean-field version of Ohm's law in the form
\begin{equation}
\mathbf{D} \cdot \overline{\mathbf{j}} = \Em + \vm \times \bm + \emf^* \, . \label{eqc:6}
\end{equation}
$\mathbf{D}$ means the magnetic diffusivity tensor (\ref{eq6:20}) and $\emf^*$ the electromotive
force $\emf$ without the $\boldsymbol{\beta}$ term, $\emf^* = \emf + \boldsymbol{\beta} \cdot
(\nabla \times \bm)$. The energy balance (\ref{eqc:4}) turns with (\ref{eqc:6}) into
\begin{equation}
\frac{d}{dt}\int_\infty \frac{\bm^2}{2\mu}dv
     = - \int_\mathcal{V} \big(D_{ij} \overline{j}_i \overline{j}_j
     - \overline{\mathbf{j}} \cdot (\vm \times \bm)
     - \overline{\mathbf{j}} \cdot \emf^* \big)\,dv \, .
\label{eqc:8}
\end{equation}
If $\mathbf{D}$ is positive definite, the first term under the right integral clearly describes
a decrease of the energy stored in the mean magnetic field. In the absence of the other two terms
the field would be bound to decay. By the way, a part of $\emf^*$ with the structure
$\boldsymbol{\delta} \times (\bnabla \times \bm)$ does not contribute to $\overline{\mathbf{j}}
\cdot \emf^*$.

We recall here that the relations (\ref{eq5:6}) have been used for the determination of
$\boldsymbol{\beta}$ and so $\mathbf{D}$ as well as $\boldsymbol{\alpha}$,
$\boldsymbol{\gamma}$, $\boldsymbol{\delta}$ and $\boldsymbol{\kappa}$ from the $\tilde a_{\kappa
\lambda}$, $\tilde b_{\kappa \lambda r}$ and $\tilde b_{\kappa \lambda \vartheta}$ extracted from
the numerical simulations, and that these relations have been derived with the particular choice
$\slb_{\kappa \lambda \varphi} = 0$. In the magnetoconvection case $\mathbf{D}$ proved to be
positive definite. Another choice for $\slb_{\kappa \lambda \varphi}$ would lead to another
$\mathbf{D}$ and another $\emf^*$. In particular, $\mathbf{D}$ can lose its definiteness.
However, the physical situation can not change by these redefinitions. $\emf^*$ also changes, such
that the right-hand side of (\ref{eqc:8}) remains its value.

By these reasons the deviations of $\mathbf{D}$ from being positive definite in the geodynamo
case do not seem to be dramatic. The growth of the magnetic energy suggested by these deviations
may well be intercepted by induction effects described by $\emf^*$. Presumably also the negative
diffusivity found in the investigations by \citet{brandenburg_02} need not to be considered as
unphysical but could also be understood in that sense.

%\begin{acknowledgments}
%\end{acknowledgments}

\bibliography{schrinner}
\bibliographystyle{gafd-eig}

\end{document}